\documentclass[11pt]{article}

\usepackage{graphicx,amsmath,amsfonts,amssymb, dcolumn,amsthm}
\usepackage{amsmath,amssymb,amsfonts}	
\usepackage{slashed}            
\usepackage{bbm}                
\usepackage{xspace}				

\numberwithin{equation}{section} 
\usepackage{tikz}
\usetikzlibrary{arrows,shapes}
\usetikzlibrary{trees}
\usetikzlibrary{matrix,arrows} 				
\usetikzlibrary{positioning}				
\usetikzlibrary{calc,through}				
\usetikzlibrary{decorations.pathreplacing}  
\usepackage{pgffor}							

\usetikzlibrary{decorations.pathmorphing}	
\usetikzlibrary{decorations.markings}
\tikzset{
    vector/.style={decorate, decoration={snake}, draw},
	provector/.style={decorate, decoration={snake,amplitude=2.5pt}, draw},
	antivector/.style={decorate, decoration={snake,amplitude=-2.5pt}, draw},
    fermion/.style={draw=black, postaction={decorate},
        decoration={markings,mark=at position .55 with {\arrow[draw=black]{latex}}}},
    fermionbar/.style={draw=black, postaction={decorate},
        decoration={markings,mark=at position .55 with {\arrow[draw=black]{latex reversed}}}},
    fermionnoarrow/.style={draw=black},
    gluon/.style={decorate, draw=black,
        decoration={coil,amplitude=4pt, segment length=5pt}},
    scalar/.style={dashed,draw=black, postaction={decorate},
        decoration={markings,mark=at position .55 with {\arrow[draw=black]{latex}}}},
    scalarbar/.style={dashed,draw=black, postaction={decorate},
        decoration={markings,mark=at position .55 with {\arrow[draw=black]{latex}}}},
    scalarnoarrow/.style={dashed,draw=black},
    electron/.style={draw=black, postaction={decorate},
        decoration={markings,mark=at position .55 with {\arrow[draw=black]{latex}}}},
	bigvector/.style={decorate, decoration={snake,amplitude=4pt}, draw},
}

\tikzstyle{block} = [draw, rectangle, 
    minimum height=3em, minimum width=6em]

\begin{document}

\title{\textbf{Remarks on the renormalization properties of Lorentz- and CPT-violating quantum electrodynamics}}
\author{ \textbf{Tiago R.~S.~Santos}\thanks{tiagoribeiro@if.uff.br}\  \ and \textbf{Rodrigo F.~Sobreiro}\thanks{sobreiro@if.uff.br}\\\\
\textit{{\small UFF $-$ Universidade Federal Fluminense,}}\\
\textit{{\small Instituto de F\'{\i}sica, Campus da Praia Vermelha,}}\\
\textit{{\small Avenida General Milton Tavares de Souza s/n, 24210-346,}}\\
\textit{{\small Niter\'oi, RJ, Brasil.}}}
\date{}
\maketitle

\begin{abstract}
In this work, we employ algebraic renormalization technique to show the renormalizability to all orders in perturbation theory of the Lorentz- and CPT-violating QED. Essentially, we control the breaking terms by using a suitable set of external sources. Thus, with the symmetries restored, a perturbative treatment can be consistently employed. After showing the renormalizability, the external sources attain certain physical values, which allow the recovering of the starting physical action. The main result is that the original QED action presents the three usual independent renormalization parameters. The Lorentz-violating sector can be renormalized by nineteen independent parameters. Moreover, vacuum divergences appear with extra independent renormalization. Remarkably, the bosonic odd sector (Chern-Simons-like term) does not renormalize and is not radiatively generated. One-loop computations are also presented and compared with the existing literature.
\end{abstract}

\section{Introduction}\label{intro}

In the last few decades many efforts have been employed in order to understand models that present Lorentz- and CPT-symmetry breaking, see for instance \cite{Colladay:1996iz,Colladay:1998fq,Jackiw:1999yp,Kostelecky:2003fs,Kostelecky:2005ic,Diaz:2011ia}. In particular, the main contributions are interested in how these models are situated under aspects of the usual quantum field theory. Due to the well known success of quantum field theory -- specially, gauge field theory -- in describing at least three of four fundamental interactions, any extension of the standard model respecting attributes as stability, renormalizability, unitarity and causality could be interesting. In fact, the Lorentz and gauge symmetry have a fundamental importance on the features mentioned before. For instance, the functional that describes the dynamics of the fields belonging to the standard model are built in a Lorentz covariant way and the classification of particles is performed by studying the Lorentz group representations \cite{Bargmann:1946me,Bargmann:1948ck}. Moreover, besides restricting the coupling between fields, the gauge symmetry play an important role on unitarity and renormalizability of gauge theories \cite{Becchi:1975nq,Tyutin:1975qk,Kugo:1979gm}.

The Abelian Lorentz- and CPT-violating minimal Standard Model Extension (mSME), \textit{i.e}., Lorentz- and CPT-violating QED, is characterized by the presence of constant background tensorial (and pseudo-tensorial) fields coupled to the fundamental fields of the theory, and is power-counting renormalizable \cite{Colladay:1998fq}. These background fields are, in principle, natural consequences of more fundamental theories such as string theories \cite{Kostelecky:1988zi}, non-commutative field theories \cite{Guralnik:2001ax, Carroll:2001ws, Carmona:2002iv, Carlson:2002zb, SheikhJabbari:2000vi}, supersymmetric field theories \cite{Berger:2003ay, Berger:2001rm, GrootNibbelink:2004za} and loop quantum gravity \cite{Alfaro:2004ur}. For instance, there exists the possibility of spontaneous Lorentz symmetry breaking in string theory. This breaking manifests itself when tensorial fields acquire non-trivial vacuum expectation values. This feature implies on a preferred spacetime direction. Although many searches have been performed in order to detect signs of these background tensors \cite{Bluhm:1997ci,Bear:2000cd,Bluhm:1999ev,Bluhm:1998rk}, nothing have been found so far. Nevertheless, these efforts have been useful to determine phenomenological and experimental upper bounds for the v.e.v. of these tensors \cite{Kostelecky:2008ts}. 

In what concerns the theoretical consistency of these models, it has been verified that they can preserve causality and unitary \cite{Colladay:1996iz,Colladay:1998fq,Adam:2001ma,Adam:2001kx,Kostelecky:2000mm,Adam:2002rg}. In this work, we confine ourselves to the formal analysis of renormalizability of the Lorentz- and CPT-violating QED. In fact, there are some works about the renormalizability of such models. For instance: in \cite{Kostelecky:2001jc}, the one-loop renormalization is discussed; the proof of renormalizability to all orders in perturbation theory, from algebraic renormalization technique point of view \cite{Piguet:1995er}, was performed in \cite{DelCima:2012gb}. The latter makes use of the gauge symmetry and requires PT-invariance to prove that anomalies are not present. Essentially, they prove the renormalizability of the model with C and/or PT-invariance. Moreover, they find nine independent renormalization parameters; furthermore, they also show in \cite{DelCima:2009ta} that no CPT-odd bosonic Lorentz violation is generated from the CPT-odd fermionic Lorentz violation sector; there also exist studies about the renormalization properties of the QED extension on curved manifolds \cite{deBerredoPeixoto:2006wz}. In this work the renormalization study was realized by assuming that Lorentz- and CPT-violating parameters are classical fields rather than constants. This last approach shares some resemblance with the present work.

It is worth mention that there also exists a class of Lorentz-violating quantum field theories which can also preserve unitarity and renormalizability, see for instance \cite{Anselmi:2007ri}. These theories are characterized by the presence of higher order space derivatives while the time derivatives remain at the same order of the usual fermionic or bosonic models. The renormalizability is assured by modifying the usual power counting criterion by introducing the concept of ``weighted power-counting" \cite{Anselmi:2014eba}.  Actually, they introduce a regulator to account for this discrepancy. This criterion can put on a renormalizable form vertexes that, in principle, are non renormalizable \cite{Anselmi:2010zh}. For instance, a higher-energy Lorentz-violating QED shows itself to be super-renormalizable and its low-energy limit is recovered by choosing specific values for scale parameters \cite{Anselmi:2009ng}. In this work, however, we deal with Carroll-Field-Jackiw theories \cite{Carroll:1989vb}, which constitute a whole different class of theories. For instance, space and time are treated on an equal footing, and the Lorentz violation manifests itself under particle Lorentz transformations. On the other hand, these theories are manifestly invariant under observer Lorentz transformations. 

In this work, we employ the BRST quantization and algebraic renormalization theory to explore the renormalizability of Lorentz- and CPT-violating QED. In particular, we generalize the study made in \cite{DelCima:2012gb} for all possible Lorentz breaking terms. Furthermore, in this work we add one more breaking term not considered in \cite{Kostelecky:2001jc}, which is a massive term coupled to a pseudo-scalar operator. The main idea can be summarized in the following way: The action which describes the bosonic Lorentz violation of CPT-odd is gauge invariant only because the Lorentz-violating coefficients are constant (and also neglecting surface terms). Thus, the gauge symmetry is ensured at the action level, but not at Lagrangian level.

The algebraic renormalization approach, which will be used here, relies on the quantum action principle (QAP) \cite{Piguet:1995er, Piguet:1980nr,Lowenstein:1971vf, Clark:1976ym, Lam:1972mb,Duetsch:2000nh}. Thus, in order to analyze the renormalizability of the Lorentz-violating QED we will employ here the Symanzik method \cite{Symanzik:1969ek} -- vastly employed in non-Abelian gauge theories in order to control a soft BRST symmetry breaking, see \cite{Zwanziger:1992qr,Dudal:2005na,Baulieu:2008fy,Baulieu:2009xr,Dudal:2011gd,Pereira:2013aza} -- to treat the BRST quantization of the Lorentz-violating electrodynamics. We will follow here the procedure employed in the proof of the renormalizability to all orders in perturbation theory of pure Yang-Mills (YM) theory with Lorentz violation \cite{Santos:2014lfa}.

The main results obtained from our approach are: First, the model is renormalizable to all orders in perturbation theory; second, the usual QED sector has only  three independent renormalization parameters (in accordance with the usual QED); third, the Chern-Simons-like violating term does not renormalize. Moreover, we attain extra important results: The Lorentz-violating sector has nineteen renormalization parameters; extra independent renormalization parameters are needed to account for extra vacuum divergences that do not appear from other approaches. However, these terms do not affect the dynamical content of the model; as pointed out in \cite{DelCima:2009ta}, the Abelian Chern-Simons-like term is not induced from the CPT-odd Lorentz-violating term of the fermionic sector, see also \cite{Jackiw:1999yp,Bonneau:2000ai,PerezVictoria:1999uh}; although one-loop computations have already been done in \cite{Kostelecky:2001jc}, we also perform these computations here in order to compare them with the algebraic results.

This work is organized as follows: In Sect.~\ref{QED} we provide the definitions, conventions and some properties of the Lorentz-violating electrodynamics. In Sect.~\ref{QUATIZATION}, the BRST quantization of the model with the extra set of auxiliary sources is provided: where we discuss the subtle quantization of the Lorentz-violating coefficients coupled with this respective composite operators within of this formalism. In Sect.~\ref{RENORMALIZABILITY}, we study the renormalizability properties of the model, with a detailed study of the quantum stability of the model. Then, in Sect.~\ref{ONELOOP}, we present the one-loop explicit computations and its relation with our renormalization independent scheme is discussed. Our final considerations are displayed in Sect.~\ref{FINAL}.

\section{Lorentz-violating electrodynamics}\label{QED}

The QED extension, just like the standard QED, is a gauge theory for the $U(1)$ group, where the electromagnetic field is coupled to the Dirac field through minimal coupling. However, this theory presents Lorentz violation in both, bosonic and fermionic, sectors. The breaking sectors are characterized by the presence of background fields. The model is described by following action \cite{Colladay:1998fq, Kostelecky:2003fs}
\begin{eqnarray}
S_{QEDex}&=&S_{QED}+S_{LV}\;,
\label{0}
\end{eqnarray}
where
\begin{eqnarray}
S_{QED}&=&\int d^4x\left\{-\frac{1}{4}F^{\mu\nu}F_{\mu\nu}+\overline{\psi}(i\gamma^{\mu}D_{\mu}-m)\psi\right\}\;,
\label{1}
\end{eqnarray}
is the classical action of the usual QED. The covariant derivative is defined as $D_{\mu}=\partial_{\mu}+ieA_{\mu}$, the field strength is written as $F_{\mu\nu}=\partial_{\mu}A_{\nu}-\partial_{\nu}A_{\mu}$ and $A_\mu$ is the gauge potential. The parameter $m$ stands for the electron mass and $e$ for its electric charge. The $\gamma^\mu$ matrices are in Dirac representation (see \cite{Itzykson:1980rh} for the full set of conventions\footnote{For self-consistency, we just define
\begin{eqnarray}
\gamma^5&=&\gamma_5=i\gamma^0\gamma^1\gamma^2\gamma^3=-\frac{i}{4!}\epsilon_{\mu\nu\alpha\beta}\gamma^{\mu}\gamma^{\nu}\gamma^{\alpha}\gamma^{\beta}\ ,\nonumber\\
\sigma^{\mu\nu}&=&\frac{i}{2}\left[\gamma^{\mu},\gamma^{\nu}\right]\;.
\label{DEF}
\end{eqnarray}}). The other term in \eqref{0} is the Lorentz-violating sector,
\begin{eqnarray}
S_{LV}&=&\int d^4x\left\{\epsilon_{\mu\nu\alpha\beta}v^{\mu}A^{\nu}\partial^{\alpha}A^{\beta}-\frac{1}{4}\kappa_{\alpha\beta\mu\nu}F^{\alpha\beta}F^{\mu\nu}+\overline{\psi}i\Gamma^{\mu}D_{\mu}\psi-\overline{\psi}M\psi\right\}\;,
\label{2}
\end{eqnarray}
where,
\begin{eqnarray}
\Gamma^{\mu}&\equiv&c^{\nu\mu}\gamma_{\nu}+d^{\nu\mu}\gamma_5\gamma_{\nu}+e^{\mu}+if^{\mu}\gamma_5+\frac{1}{2}g^{\alpha\beta\mu}\sigma_{\alpha\beta}\ ,\nonumber\\
M&\equiv&im_5\gamma_5+a^{\mu}\gamma_{\mu}+b^{\mu}\gamma_5\gamma_{\mu}+\frac{1}{2}h^{\mu\nu}\sigma_{\mu\nu}\;.
\label{3}
\end{eqnarray}
The violation of Lorentz symmetry in the fermionic sector is characterized by the following constant tensorial fields: $c^{\nu\mu}$, $d^{\nu\mu}$, $e^{\mu}$, $f^{\mu}$, $g^{\alpha\beta\mu}$, $m_5$, $a^{\mu}$, $b^{\mu}$ and $h^{\mu\nu}$. These tensors select privileged directions in spacetime, dooming it to anisotropy. Tensorial fields with even numbers of indexes preserve CPT while an odd number of indexes do not preserve CPT\footnote{For the explicit CPT features of the fields see Table \ref{table4} in terms of the sources at the App.~\ref{DM} (See next section for the source-background correspondence).}. The tensorial fields $c^{\nu\mu}$, $d^{\nu\mu}$, $e^{\mu}$, $f^{\mu}$ and $g^{\alpha\beta\mu}$ are dimensionless and $m_5$, $a^{\mu}$, $b^{\mu}$ and $h^{\mu\nu}$ has mass dimension 1.  The tensor field $h^{\mu\nu}$ is anti-symmetric and $g^{\alpha\beta\mu}$ is anti-symmetric only on its first two indexes. At the photonic sector, the Lorentz violation is characterized by the field $v^{\mu}$, with mass dimension 1, and $\kappa_{\alpha\beta\mu\nu}$, which is dimensionless. The tensor $\kappa_{\alpha\beta\mu\nu}$ obeys the same properties of the Riemann tensor, and is double traceless
\begin{align}
&\kappa_{\alpha\beta\mu\nu}\;=\;\kappa_{\mu\nu\alpha\beta}\;=\;-\kappa_{\beta\alpha\mu\nu}\ ,\nonumber\\
&\kappa_{\alpha\beta\mu\nu}+\kappa_{\alpha\mu\nu\beta}+\kappa_{\alpha\nu\beta\mu}\;=\;0\ ,\nonumber\\
&\kappa^{\mu\nu}_{\phantom{\mu\nu}\mu\nu}\;=\;0\ .
\label{3aaa}
\end{align}
As the reader can easily infer, the action \eqref{0} is a Lorentz scalar, being invariant under observers Lorentz transformations while, in contrast, presents violation with respect to particle Lorentz transformations \footnote{This can be understood in the following way: let $\Phi^\mu$ and $Q^\mu$ be a generic field and a background vector field, respectively. Under observer Lorentz transformation, \textit{i.e.}, exchange of references systems, these fields behave as $\Phi'^\mu =\Lambda^{\mu}_{\phantom{\mu}\nu}\Phi^\nu $ and $Q'^\mu =\Lambda^{\mu}_{\phantom{\mu}\nu}Q^\nu$. On the other hand, under particle Lorentz transformation the reference systems do not transform, but the fields transform as $\Phi'^\mu =\Lambda^{\mu}_{\phantom{\mu}\nu}\Phi^\nu $ and $Q'^\mu =Q^\mu $. The generalization to (pseudo-)tensorial backgrounds are immediate.} \cite{Kostelecky:2001jc}.

\section{BRST quantization and restoration of Lorentz symmetry} \label{QUATIZATION}

In the process of quantization of the QED extension theory, as in the usual QED, gauge fixing is required. In the present work we employ the BRST quantization method and adopt, for simplicity, the Landau gauge condition $\partial_\mu A^{\mu}=0$. Thus, besides the photon and electron fields, we introduce the Lautrup-Nakanishi field $b$ and the Faddeev-Popov ghost and anti-ghost fields\footnote{Even though the ghost and anti-ghost fields are not required in the Abelian theory at Landau gauge, we opt by introduce them for following reasons: i) It is a direct way to keep the off-shell BRST symmetry of the action $S_0$. ii) Due to the discrete Faddeev-Popov symmetry, the trivial and non-trivial sectors of the BRST cohomology becomes explicit (see table \ref{table1}). iii) With the introduction of the $b$ field, it is easy to see that photon propagator keeps its transversality to all orders in perturbation theory. iv) As expected, the tree-level decoupling of the ghosts is kept at all orders in perturbation theory, see ~Sect. \ref{WI}. This feature will bring important consequences for the renormalization properties of the CPT-odd bosonic violating sector of the action \eqref{0}.}, namely, $c$ and $\overline{c}$, respectively. The BRST transformations are
\begin{eqnarray}
sA_{\mu}&=&-\partial_{\mu}c\;,\nonumber\\
sc&=&0\;,\nonumber\\
s\psi&=&iec\psi\;,\nonumber\\
s\overline{\psi}&=&ie\overline{\psi}c\;,\nonumber\\
s\overline{c}&=&b\;,\nonumber\\
sb&=&0\;,\label{4a}
\end{eqnarray}
where $s$ is the nilpotent BRST operator. Thus, the Landau gauge fixed action is
\begin{eqnarray}
S_0&=&S_{QED}+S_{LV}+S_{gf}\;,
\label{4b}
\end{eqnarray}
where 
\begin{eqnarray}
S_{gf}&=&s\int d^4x\overline{c}\partial_{\mu}A^{\mu}=\int d^4x\left(b\partial_{\mu}A^{\mu}+\overline{c}\partial^2c\right)\ ,
\label{6}
\end{eqnarray}
is the gauge fixing action enforcing the Landau gauge condition. The quantum numbers of the fields and background tensors are displayed in tables \ref{table1} and \ref{table2}, respectively.

\begin{table}[h]
\centering
\begin{tabular}{|c|c|c|c|c|c|c|}
	\hline 
fields & $A$ & $b$ &$c$ & $\overline{c}$ & $\psi$ & $\overline{\psi}$ \\
	\hline 
UV dimension & $1$ & $2$ & $0$ & $2$ & $3/2$& $3/2$\\ 
Ghost number & $0$ & $0$ & $1$& $-1$& $0$& $0$\\ 
Spinor number & $0$ & $0$ & $0$& $0$& $1$&$-1 $\\ 
Statistics & $0$& $0$ & $1$ & $-1$& $1$ & $-1$ \\ 
\hline 
\end{tabular}
\caption{Quantum numbers of the fields.}
\label{table1}
\end{table}

\begin{table}[h]
\centering
\begin{tabular}{|c|c|c|c|c|c|c|c|c|c|c|c|}
	\hline 
tensors & $v$ & $\kappa$ & $c$ & $d$ & $e$ & $f$ & $g$ & $m_5$ & $a$ & $b$ & $h$ \\
	\hline 
UV dimension & $1$ & $0$ & $0$ & $0$ & $0$ & $0$ & $0$ & $1$ & $1$ & $1$ & $1$ \\ 
Ghost number & $0$ & $0$ & $0$ & $0$ & $0$ & $0$ & $0$ & $0$ & $0$ & $0$ & $0$ \\ 
Spinor number & $0$ & $0$ & $0$ & $0$ & $0$ & $0$ & $0$ & $0$ & $0$ & $0$ & $0$  \\ 
Statistics & $0$ & $0$  & $0$ & $0$  & $0$ & $0$ & $0$ & $0$ & $0$ & $0$ & $0$ \\ 
\hline 
\end{tabular}
\caption{Quantum numbers of the background tensors.}
\label{table2}
\end{table}

Following the BRST quantization of the Lorentz-violating sector with the Symanzik prescription \cite{Santos:2014lfa}, we will have two distincts situations. The first situation concerns the CPT-even bosonic violating term and all fermionic breaking terms, all of them are BRST invariant. Then, they will couple to BRST invariant sources. Henceforth, we define the following set of invariant sources
\begin{eqnarray}
s\bar{\kappa}_{\alpha\beta\mu\nu}=sC^{\nu\mu}=sD^{\nu\mu}=sE^{\mu}=sF^{\mu}=sG^{\alpha\beta\mu}=sM_5=s\bar{A}^{\mu}=sB^{\mu}=sH^{\mu\nu}=0\;.
\label{10}
\end{eqnarray}
On the second situation, the CPT-odd bosonic violating term, a BRST doublet is required because this term is not BRST invariant,
\begin{eqnarray}
s\lambda_{\mu\nu\alpha}&=&J_{\mu\nu\alpha}\;, \nonumber\\
sJ_{\mu\nu\alpha}&=&0\;.
\label{9}
\end{eqnarray}
The quantum numbers of the sources are displayed in table \ref{table3}. Eventually, in order to re-obtain the starting action \eqref{4b}, these sources will attain the following physical values
\begin{eqnarray}
J_{\mu\nu\alpha}\mid_{phys}&=&v^{\beta}\epsilon_{\beta\mu\nu\alpha}\;,\nonumber\\
\lambda_{\mu\nu\alpha}\mid_{phys}&=&0\ ,\nonumber\\
\bar{\kappa}_{\alpha\beta\mu\nu}\mid_{phys}&=&\kappa_{\alpha\beta\mu\nu}\;,\nonumber\\
C^{\nu\mu}\mid_{phys}&=&c^{\nu\mu}\;,\nonumber\\
D^{\nu\mu}\mid_{phys}&=&d^{\nu\mu}\;,\nonumber\\
E^{\mu}\mid_{phys}&=&e^{\mu}\;,\nonumber\\
F^{\mu}\mid_{phys}&=&f^{\mu}\;,\nonumber\\
G^{\alpha\beta\mu}\mid_{phys}&=&g^{\alpha\beta\mu}\;,\nonumber\\
M_5\mid_{phys}&=&m_5\;,\nonumber\\
\bar{A}^{\mu}\mid_{phys}&=&a^{\mu}\ ,\nonumber\\
B^{\mu}\mid_{phys}&=&b^{\mu}\;,\nonumber\\
H^{\mu\nu}\mid_{phys}&=&h^{\mu\nu}\;.
\label{12}
\end{eqnarray}

\begin{table}[h]
\centering
\begin{tabular}{|c|c|c|c|c|c|c|c|c|c|c|c|c|c|c|}
	\hline 
sources & $Y$ & $\overline{Y}$ &$\lambda$ & $J$ & $\bar{\kappa}$ & $C$ & $D$ & $E$ & $F$ & $G$ & $M_5$ & $\bar{A}$ & $B$ & $H$ \\
	\hline 
UV dimension & $5/2$ & $5/2$ & $1$ & $1$ & $0$ & $0$ & $0$ & $0$ & $0$ & $0$ & $1$ & $1$ & $1$ & $1$ \\ 
Ghost number & $-1$ & $-1$ & $-1$ & $0$ & $0$ & $0$ & $0$ & $0$ & $0$ & $0$ & $0$ & $0$ & $0$ & $0$ \\ 
Spinor number & $1$ & $-1$ & $0$ & $0$ & $0$ & $0$ & $0$ & $0$ & $0$ & $0$ & $0$ & $0$ & $0$ & $0$  \\ 
Statistics &  $0$ & $-2$ & $-1$ & $0$ & $0$  & $0$ & $0$  & $0$ & $0$ & $0$ & $0$ & $0$ & $0$ & $0$ \\ 
\hline 
\end{tabular}
\caption{Quantum numbers of the sources.}
\label{table3}
\end{table}

Thus, we replace the action \eqref{4b} by 
\begin{eqnarray}
S&=&S_{QED}+S_{B}+S_{F}+S_{gf}\;,
\label{5}
\end{eqnarray}
where 
\begin{eqnarray}
S_{B}&=&s\int d^4x\lambda_{\mu\nu\alpha}A^{\mu}\partial^{\nu}A^{\alpha}-\frac{1}{4}\int d^4x\;\bar{\kappa}_{\alpha\beta\mu\nu}F^{\alpha\beta}F^{\mu\nu}\;,\nonumber\\
&=&\int d^4x\left(J_{\mu\nu\alpha}A^{\mu}\partial^{\nu}A^{\alpha}+\lambda_{\mu\nu\alpha}\partial^{\mu}c\partial^{\nu}A^{\alpha}\right)-\frac{1}{4}\int d^4x\;\bar{\kappa}_{\alpha\beta\mu\nu}F^{\alpha\beta}F^{\mu\nu}\;,
\label{7}
\end{eqnarray}
is the embedding\footnote{The embedding concept used here is discussed in detail in Ref.~\cite{Santos:2014lfa}.} of the Lorentz-violating bosonic sector while the embedding of the Lorentz-violating term for the fermionic sector is given by
\begin{eqnarray}
S_{F}&=&\int d^4x\left\{i\left(C^{\nu\mu}\overline{\psi}\gamma_{\nu}D_{\mu}\psi+D^{\nu\mu}\overline{\psi}\gamma_{5}\gamma_{\nu}D_{\mu}\psi+E^{\mu}\overline{\psi}D_{\mu}\psi+iF^{\mu}\overline{\psi}\gamma_{5}D_{\mu}\psi+\right.\right.\nonumber\\
&+&\left.\left.\frac{1}{2}G^{\alpha\beta\mu}\overline{\psi}\sigma_{\alpha\beta}D_{\mu}\psi\right)-\left(iM_5\overline{\psi}\gamma_5\psi+\bar{A}^{\mu}\overline{\psi}\gamma_{\mu}\psi+B^{\mu}\overline{\psi}\gamma_5\gamma_{\mu}\psi+\frac{1}{2}H^{\mu\nu}\overline{\psi}\sigma_{\mu\nu}\psi\right)\right\}\;.\nonumber\\
\end{eqnarray}
It is easy to check that $s\left(S_{QED}+S_{B}+S_{F}+S_{gf}\right)=0$. The quantum number of the sources follow the quantum numbers of the background fields, as displayed in\footnote{The external sources $\overline{Y}$ and $Y$ will be defined in Section \ref{RENORMALIZABILITY} in order to control the nonlinear BRST transformations of the spinor fields.} Table \ref{table3}.

The action $S$, at the physical value of the sources \eqref{12}, reduces to
\begin{eqnarray}
\Sigma_{phys}&=&\int d^4x\left\{-\frac{1}{4}F^{\mu\nu}F_{\mu\nu}+\overline{\psi}(i\gamma^{\mu}D_{\mu}-m)\psi\right\}+\int d^4x\left(b\partial_{\mu}A^{\mu}+\overline{c}\partial^2c\right)+\nonumber\\
&+&\int d^4x\left(\epsilon_{\beta\mu\nu\alpha}v^\beta A^{\mu}\partial^{\nu}A^{\alpha}-\frac{1}{4}{\kappa}_{\alpha\beta\mu\nu}F^{\alpha\beta}F^{\mu\nu}\right)+\nonumber\\
&+&\int d^4x\left\{i\left(c^{\nu\mu}\overline{\psi}\gamma_{\nu}D_{\mu}\psi+d^{\nu\mu}\overline{\psi}\gamma_{5}\gamma_{\nu}D_{\mu}\psi+e^{\mu}\overline{\psi}D_{\mu}\psi+if^{\mu}\overline{\psi}\gamma_{5}D_{\mu}\psi+\right.\right.\nonumber\\
&+&\left.\left.\frac{1}{2}g^{\alpha\beta\mu}\overline{\psi}\sigma_{\alpha\beta}D_{\mu}\psi\right)-\left(im_5\overline{\psi}\gamma_5\psi+a^{\mu}\overline{\psi}\gamma_{\mu}\psi+b^{\mu}\overline{\psi}\gamma_5\gamma_{\mu}\psi+\frac{1}{2}h^{\mu\nu}\overline{\psi}\sigma_{\mu\nu}\psi\right)\right\}\:.\nonumber\\
\label{15aab}
\end{eqnarray}

It is clear then that the kinematical content of the model does not change once the physical limit of the sources are taken. This is a peculiarity of the Abelian model, where the symmetries avoid many terms that are present at non-Abelian model \cite{Santos:2014lfa}. In fact, at the non-Abelian model with Lorentz violation the kinematics of the model is drastically changed when this approach is employed. See \cite{Santos:2014lfa} for more details.

\section{Algebraic proof of the renormalizability}\label{RENORMALIZABILITY}

Let us now face the issue of the renormalizability of the model. For that, we need one last set of external BRST invariant sources, namely, $\overline{Y}$ and $Y$, in order to control the non-linear BRST transformations of the original fields,
\begin{eqnarray}
S_{ext}&=&\int d^4x\left(\overline{Y}s\psi-s\overline{\psi}Y\right)=\int d^4x\left(ie\overline{Y}c\psi-ie\overline{\psi}cY\right)\;.
\label{13}
\end{eqnarray}
Thus, the complete action is given by
\begin{eqnarray}
\Sigma&=&S+S_{ext}\:.
\label{15}
\end{eqnarray}
Indeed, it is easy to note that extra combinations among sources are possible, including the electron mass. However, these combinations do not interfere with the renormalization of the sources and they will be renormalizable as well. Moreover, extra dimensionless parameters will be needed in order to absorb vacuum divergences, see for instance \cite{Santos:2014lfa}. To avoid a cumbersome analysis, we omit these pure vacuum terms here. Nevertheless, for completeness, this issue is discussed in the Appendix \ref{Vacuumterms}.

Explicitly, the action \eqref{15} has the form  
\begin{eqnarray}
\Sigma&=&\int d^4x\left\{-\frac{1}{4}F^{\mu\nu}F_{\mu\nu}+\overline{\psi}(i\gamma^{\mu}D_{\mu}-m)\psi\right\}+\int d^4x\left(b\partial_{\mu}A^{\mu}+\overline{c}\partial^2c\right)+\nonumber\\
&+&\int d^4x\left(J_{\mu\nu\alpha}A^{\mu}\partial^{\nu}A^{\alpha}+\lambda_{\mu\nu\alpha}\partial^{\mu}c\partial^{\nu}A^{\alpha}-\frac{1}{4}\bar{\kappa}_{\alpha\beta\mu\nu}F^{\alpha\beta}F^{\mu\nu}\right)+\nonumber\\
&+&\int d^4x\left\{i\left(C^{\nu\mu}\overline{\psi}\gamma_{\nu}D_{\mu}\psi+D^{\nu\mu}\overline{\psi}\gamma_{5}\gamma_{\nu}D_{\mu}\psi+E^{\mu}\overline{\psi}D_{\mu}\psi+iF^{\mu}\overline{\psi}\gamma_{5}D_{\mu}\psi+\right.\right.\nonumber\\
&+&\left.\left.\frac{1}{2}G^{\alpha\beta\mu}\overline{\psi}\sigma_{\alpha\beta}D_{\mu}\psi\right)-\left(iM_5\overline{\psi}\gamma_5\psi+\bar{A}^{\mu}\overline{\psi}\gamma_{\mu}\psi+B^{\mu}\overline{\psi}\gamma_5\gamma_{\mu}\psi+\frac{1}{2}H^{\mu\nu}\overline{\psi}\sigma_{\mu\nu}\psi\right)\right\}+\nonumber\\
&+&\int d^4x\left(ie\overline{Y}c\psi-ie\overline{\psi}cY\right)\:.
\label{15aaab}
\end{eqnarray}
As one can easily check that, at the physical values of the sources, this action is also contracted down to \eqref{15aab}.

\subsection{Ward identities}\label{WI}

The action \eqref{15} enjoys the following set of Ward identities

\begin{itemize}
	\item Slavnov-Taylor identity
	\begin{eqnarray}
\mathcal{S}(\Sigma)&=&\int d^4x\left(-\partial_{\mu} c\frac{\delta \Sigma}{\delta A_{\mu}}+\frac{\delta \Sigma}{\delta \overline{Y}}\frac{\delta \Sigma}{\delta \psi}-\frac{\delta \Sigma}{\delta Y}\frac{\delta \Sigma}{\delta \overline{\psi}}+b\frac{\delta \Sigma}{\delta \overline{c}}+J_{\mu\nu\alpha}\frac{\delta \Sigma}{\delta\lambda_{\mu\nu\alpha}}\right)=0\;.
\label{16}
\end{eqnarray}

\item Gauge fixing and anti-ghost equations
\begin{eqnarray}
\frac{\delta \Sigma}{\delta b}&=&\partial_{\mu}A^{\mu}\;,\nonumber\\
\frac{\delta \Sigma}{\delta \overline{c}}&=&\partial^2c\;.
\label{17}
\end{eqnarray}

\item Ghost equation
\begin{eqnarray}
\frac{\delta \Sigma}{\delta c}&=&\partial^\mu\left(\lambda_{\mu\nu\alpha}\partial^\nu A^\alpha\right)-\partial^2\overline{c}+ie\overline{Y}\psi+ie\overline{\psi}Y\;.
\label{19}
\end{eqnarray}
\end{itemize}
At \eqref{17} and \eqref{19}, the breaking terms are linear in the fields. Thus, they will remain at classical level \cite{Piguet:1995er}, a property that is guaranteed by the quantum action principle \cite{Lowenstein:1971vf}. 

\subsection{Most general counterterm}\label{MGC}

In order to obtain the most general counterterm which can be freely added to the classical action $\Sigma$ at any order in perturbation theory, we need a general local integrated polynomial $\Sigma^c$ with dimension bounded by four and vanishing ghost number. Thus, imposing the Ward identities \eqref{16}-\eqref{19} to the perturbed action $\Sigma+\varepsilon\Sigma^c$, where $\varepsilon$ is a small parameter, it is easy to find that the counterterm must obey the following constraints
\begin{eqnarray}
\mathcal{S}_{\Sigma}\Sigma^c&=&0\;,\nonumber\\
\frac{\delta \Sigma^c}{\delta b}&=&0\;,\nonumber\\
\frac{\delta \Sigma^c}{\delta \overline{c}}&=&0\;,\nonumber\\
\frac{\delta \Sigma^c}{\delta c}&=&0\;,
\label{22}
\end{eqnarray}
where the operator $\mathcal{S}_{\Sigma}$ is the nilpotent linearized Slavnov-Taylor operator,
\begin{eqnarray}
\mathcal{S}_{\Sigma}=\int d^4x\left(-\partial_{\mu} c\frac{\delta}{\delta A_{\mu}}+\frac{\delta \Sigma}{\delta \overline{Y}}\frac{\delta }{\delta \psi}+\frac{\delta \Sigma}{\delta\psi}\frac{\delta}{\delta\overline{Y}}-\frac{\delta \Sigma}{\delta Y}\frac{\delta}{\delta \overline{\psi}}-\frac{\delta \Sigma}{\delta\overline{\psi}}\frac{\delta}{\delta Y}+b\frac{\delta}{\delta \overline{c}}+J_{\mu\nu\alpha}\frac{\delta }{\delta\lambda_{\mu\nu\alpha}}\right)\;.
\label{26}
\end{eqnarray}

The first constraint  of \eqref{22} identifies the invariant counterterm as the solution of the cohomology problem for the operator $\mathcal{S}_{\Sigma}$ in the space of the integrated local field polynomials of dimension four. From the general results of cohomology, it follows that $\Sigma^c$ can be written as \cite{Piguet:1995er}
\begin{eqnarray}
\Sigma^c&=&-\frac{1}{4}\int d^4x\;a_0F^{\mu\nu}F_{\mu\nu}-\frac{1}{4}\int d^4x\left(a_1\bar{\kappa}_{\alpha\mu\beta\nu}+a_2T_{\alpha\mu\beta\nu}^{\phantom{\alpha\mu\beta\nu}\theta\omega}C_{\theta\omega}\right)F^{\alpha\mu}F^{\beta\nu}+\nonumber\\
&+&\int d^4x\left\{a_3i\overline{\psi}\gamma^{\mu}D_{\mu}\psi-a_4m\overline{\psi}\psi+ i\left[(a_5C^{\nu\mu}+ a_6C^{\mu\nu}+a_7\eta_{\alpha\beta}\bar{\kappa}^{\alpha\mu\beta\nu})\overline{\psi}\gamma_{\nu}D_{\mu}\psi+\right.\right.\nonumber\\
&+&\left.\left.\left(a_8D^{\nu\mu}+a_9D^{\mu\nu}\right)\overline{\psi}\gamma_{5}\gamma_{\nu}D_{\mu}\psi+a_{10}E^{\mu}\overline{\psi}D_{\mu}\psi+a_{11}iF^{\mu}\overline{\psi}\gamma_{5}D_{\mu}\psi+\right.\right.\nonumber\\
&+&\left.\left. \frac{1}{2}\left(a_{12}G^{\alpha\beta\gamma }+a_{13}S^{\alpha\beta\gamma}_{\phantom{\alpha\beta\gamma}\lambda\rho\sigma}G^{\lambda\rho\sigma}\right)\overline{\psi}\sigma_{\alpha\beta}D_{\gamma}\psi\right]-\left(a_{14}iM_5\overline{\psi}\gamma_5\psi+\right.\right.\nonumber\\
&+&\left.\left.(a_{15}\bar{A}^{\mu}+a_{16} mE^{\mu})\overline{\psi}\gamma_{\mu}\psi+(a_{17}B^{\mu}+a_{18}mG_{\alpha\beta\gamma}\epsilon^{\alpha\beta\gamma\mu})\overline{\psi}\gamma_5\gamma_{\mu}\psi+\right.\right.\nonumber\\
&+&\left.\left.
\frac{1}{2}(a_{19}H^{\mu\nu}+a_{20}mD_{\alpha\beta}\epsilon^{\alpha\beta\mu\nu})\overline{\psi}\sigma_{\mu\nu}\psi\right)\right\}+\mathcal{S}_{\Sigma}\Delta^{(-1)}\;,
\label{27}
\end{eqnarray}
where $\Delta^{(-1)}$ is the most general local polynomial counterterm with dimension bounded by four and ghost number $-1$, given by\footnote{Clearly, in contrast to the background fields, the external sources are not ``frozen'' with respect to CPT symmetries. Hence, they enjoy discrete mappings. Thus $\Delta^{(-1)}$ can be made CPT-invariant, avoiding a lot of counterterms and renormalization parameters. Furthermore, all of them, if included, are also avoided by the Ward identities of the model.}
\begin{eqnarray}
\Delta^{(-1)}&=&\int d^4x\left(a_{21}\overline{Y}\psi+a_{22}\overline{\psi}Y+a_{23}\overline{c}\partial_{\mu}A^{\mu}+a_{24}\overline{c}b+a_{25}\lambda_{\mu\nu\alpha}A^{\mu}\partial^{\nu}A^{\alpha}+\right.\nonumber\\
&+&\left.a_{26}\lambda_{\alpha\beta\gamma}\epsilon^{\alpha\beta\gamma\mu}\overline{\psi}\gamma_5\gamma_{\mu}\psi+a_{27}\bar{c}\partial_\mu \bar{A}^\mu+a_{28}m\bar{c}\partial_\mu E^\mu+a_{29}\lambda_{\mu\nu\alpha}J^{\mu\nu\alpha}\bar{c}c+\right.\nonumber\\ 
&+&\left.a_{30}\lambda_{\mu\nu\alpha}J^{\mu\nu\alpha}A_\beta A^\beta+a_{31}\lambda_{\mu\alpha\beta}J^{\nu\alpha\beta}A^{\mu}A_{\nu}+\right.\nonumber\\
&+&\left.a_{32}\bar{\kappa}_{\alpha\beta\mu\nu}\lambda^{\alpha\beta\rho}J^{\mu\nu}_{\phantom{\mu\nu}\rho}A_{\sigma}A^{\sigma}+a_{33}T_{\alpha\beta\mu\nu}^{\phantom{\alpha\beta\mu\nu}\theta\omega}C_{\theta\omega}\lambda^{\alpha\beta\rho}J^{\mu\nu}_{\phantom{\mu\nu}\rho}A_{\sigma}A^{\sigma}+\right.\nonumber\\
&+&\left.a_{34}\bar{\kappa}_{\alpha\beta\mu\nu}\lambda^{\beta\rho\sigma}J^{\nu}_{\phantom{\nu}\rho\sigma}A^{\alpha}A^{\mu}+a_{35}T_{\alpha\beta\mu\nu}^{\phantom{\alpha\beta\mu\nu}\theta\omega}C_{\theta\omega}\lambda^{\beta\rho\sigma}J^{\nu}_{\phantom{\nu}\rho\sigma}A^{\alpha}A^{\mu}+\right.\nonumber\\
&+&\left.a_{36}\bar{\kappa}_{\alpha\rho\sigma\delta}\lambda^{\nu\rho\delta}J^{\mu\alpha\sigma}A_{\mu}A_{\nu}+a_{37}T_{\alpha\rho\sigma\delta}^{\phantom{\alpha\rho\sigma\delta}\theta\omega}C_{\theta\omega}\lambda^{\nu\rho\delta}J^{\mu\alpha\sigma}A_{\mu}A_{\nu}\right)\;,
\label{28}
\end{eqnarray}
with $a_i$ being free coefficients, and
\begin{eqnarray}
T_{\alpha\mu\beta\nu}^{\phantom{\alpha\mu\beta\nu}\theta\omega}&\equiv&\eta_{\alpha\nu}(\delta^{\theta}_{\mu}\delta^{\omega}_{\beta}+\delta^{\theta}_{\beta}\delta^{\omega}_{\mu})-\eta_{\alpha\beta}(\delta^{\theta}_{\mu}\delta^{\omega}_{\nu}+\delta^{\theta}_{\nu}\delta^{\omega}_{\mu})-\eta_{\mu\nu}(\delta^{\theta}_{\alpha}\delta^{\omega}_{\beta}+\delta^{\theta}_{\beta}\delta^{\omega}_{\alpha})+\eta_{\mu\beta}(\delta^{\theta}_{\alpha}\delta^{\omega}_{\nu}+\delta^{\theta}_{\nu}\delta^{\omega}_{\alpha})\;,\nonumber\\
S^{\alpha\beta\gamma}_{\phantom{\alpha\beta\gamma}\lambda\rho\sigma}&\equiv&\delta^{\gamma}_{\lambda}\delta^{\beta}_{\rho}\delta^{\alpha}_{\sigma}-\delta^{\gamma}_{\lambda}\delta^{\alpha}_{\rho}\delta^{\beta}_{\sigma}+\delta^{\beta}_{\lambda}\eta_{\rho\sigma}\eta^{\alpha\gamma}-\delta^{\alpha}_{\lambda}\eta_{\rho\sigma}\eta^{\beta\gamma}\;.
\end{eqnarray}
The contraction $T_{\alpha\mu\beta\nu}^{\phantom{\alpha\mu\beta\nu}\theta\omega}C_{\theta\omega}$ has the same symmetries as the source $\bar{\kappa}_{\alpha\mu\beta\nu}$. Moreover, $S^{\alpha\beta\gamma}_{\phantom{\alpha\beta\gamma}\lambda\rho\sigma}G^{\lambda\rho\sigma}$ has the same discrete symmetries as $G^{\alpha\beta\gamma}$. Indeed, this contraction extends the anti-symmetrization to all indexes of $G^{\lambda\rho\sigma}$. From the second equation in \eqref{22}, it follows that $a_{23}=a_{24}=a_{27}=a_{28}=a_{29}=0$. Moreover, from the ghost equation, $a_{25}=a_{30}=a_{31}=a_{32}=a_{33}=a_{34}=a_{35}=a_{36}=a_{37}=0$. Then, after the following redefinitions,
\begin{align}
a_3-a_{21}+a_{22}&\mapsto a_3\;,\;\;\;\;\;\;\;\;\;a_{12}-a_{21}+a_{22}\mapsto a_{12}\;,\nonumber\\
a_4-a_{21}+a_{22}&\mapsto a_4\;,\;\;\;\;\;\;\;\;\;a_{14}-a_{21}+a_{22}\mapsto a_{14}\;,\nonumber\\
a_5-a_{21}+a_{22}&\mapsto a_5\;,\;\;\;\;\;\;\;\;\;a_{15}-a_{21}+a_{22}\mapsto a_{15}\;,\nonumber\\
a_8-a_{21}+a_{22}&\mapsto a_8\;,\;\;\;\;\;\;\;\;\;a_{17}-a_{21}+a_{22}\mapsto a_{17}\;,\nonumber\\
a_{10}-a_{21}+a_{22}&\mapsto a_{10}\;,\;\;\;\;\;\;\;\;a_{19}-a_{21}+a_{22}\mapsto a_{19}\;,\nonumber\\
a_{11}-a_{21}+a_{22}&\mapsto a_{11}\;,
\label{30}
\end{align}
it is not difficult to verify that the form of the most general counterterm allowed by the Ward identities is given by
\begin{eqnarray}
\Sigma^c&=&-\frac{1}{4}\int d^4x\;a_0F^{\mu\nu}F_{\mu\nu}-\frac{1}{4}\int d^4x\left(a_1\bar{\kappa}_{\alpha\mu\beta\nu}+a_2T_{\alpha\mu\beta\nu}^{\phantom{\alpha\mu\beta\nu}\theta\omega}C_{\theta\omega}\right)F^{\alpha\mu}F^{\beta\nu}+\nonumber\\
&+&\int d^4x\left\{a_3i\overline{\psi}\gamma^{\mu}D_{\mu}\psi-a_4m\overline{\psi}\psi+ i\left[(a_5C^{\nu\mu}+ a_6C^{\mu\nu}+a_7\eta_{\alpha\beta}\bar{\kappa}^{\alpha\mu\beta\nu})\overline{\psi}\gamma_{\nu}D_{\mu}\psi+\right.\right.\nonumber\\
&+&\left.\left.\left(a_8D^{\nu\mu}+a_9D^{\mu\nu}\right)\overline{\psi}\gamma_{5}\gamma_{\nu}D_{\mu}\psi+a_{10}E^{\mu}\overline{\psi}D_{\mu}\psi+a_{11}iF^{\mu}\overline{\psi}\gamma_{5}D_{\mu}\psi+\right.\right.\nonumber\\
&+&\left.\left. \frac{1}{2}\left(a_{12}G^{\alpha\beta\gamma }+a_{13}S^{\alpha\beta\gamma}_{\phantom{\alpha\beta\gamma}\lambda\rho\sigma}G^{\lambda\rho\sigma}\right)\overline{\psi}\sigma_{\alpha\beta}D_{\gamma}\psi\right]-\left(a_{14}iM_5\overline{\psi}\gamma_5\psi+\right.\right.\nonumber\\
&+&\left.\left.(a_{15}\bar{A}^{\mu}+a_{16} mE^{\mu})\overline{\psi}\gamma_{\mu}\psi+(a_{17}B^{\mu}+a_{18}mG_{\alpha\beta\gamma}\epsilon^{\alpha\beta\gamma\mu}-a_{26}J_{\alpha\beta\gamma}\epsilon^{\alpha\beta\gamma\mu})\overline{\psi}\gamma_5\gamma_{\mu}\psi+\right.\right.\nonumber\\
&+&\left.\left.
\frac{1}{2}(a_{19}H^{\mu\nu}+a_{20}mD_{\alpha\beta}\epsilon^{\alpha\beta\mu\nu})\overline{\psi}\sigma_{\mu\nu}\psi\right)\right\}\;.
\label{31}
\end{eqnarray}

\subsection{Stability}\label{ST}

It remains to infer if the counterterm $\Sigma^c$ can be reabsorbed by the original action $\Sigma$ by means of the multiplicative redefinition of the fields, sources and parameters of the theory, according to
\begin{eqnarray}
\Sigma(\Phi, J, \xi)+\varepsilon\Sigma^c(\Phi, J, \xi)&=&\Sigma(\Phi_0, J_0, \xi_0)+\mathcal{O}(\varepsilon^2)\;,
\label{32}
\end{eqnarray}
where the bare quantities are defined as
\begin{eqnarray}
\Phi_0&=&Z^{1/2}_{\Phi}\Phi\;,\;\;\;\;\Phi \in \left\{A, \overline{\psi},\psi,b,\overline{c},c\right\}\;,\nonumber\\
J_0&=&Z_JJ\;,\;\;\;\;\;\;\;J \in \left\{J,\lambda,C,D,E,F,G,M_5,\bar{A},B,H\right\}\;, \nonumber\\
\xi_0&=&Z_{\xi}\xi\;,\;\;\;\;\;\;\;\;\xi \in \left\{e,m\right\}\;.
\label{33}
\end{eqnarray}
It is straightforward to check that this can be performed, proving the theory to be renormalizable to all orders in perturbation theory. Explicitly, the renormalization factors are listed below. 

For the independent renormalization factors of the photon, electron and electron mass, we have
\begin{eqnarray}
Z_A^{1/2}&=&1+\frac{1}{2}\varepsilon a_0\;,\nonumber\\
Z_\psi^{1/2}&=&1+\frac{1}{2}\varepsilon a_3\;,\nonumber\\
Z_m&=&1+\varepsilon(a_4-a_3)\;.
\label{ren1}
\end{eqnarray}
The renormalization factors of the ghosts, charge, Lautrup-Nakanishi field and $Y$ sources are not independent, namely
\begin{eqnarray}
Z_c^{1/2}&=&Z_{\overline{c}}^{1/2}\;\;=\;\;1\;,\nonumber\\
Z_b^{1/2}&=&Z_e\;\;=\;\;Z_A^{-1/2}\;,\nonumber\\
Z_Y&=&Z_{\overline{Y}}\;\;=\;\;Z_A^{1/2}Z_{\psi}^{-1/2}\;.
\label{ren2}
\end{eqnarray}
Thus, the renormalization properties of the standard QED sector remain unchanged.

For the $\kappa_{\alpha\mu\beta\nu}$ sector, due to the quantum numbers of $\bar{\kappa}$ and $C$, there is a mixing between their respective operators, \emph{i.e.}, $F^{\alpha\mu} F^{\beta\nu}$ and $i\overline{\psi}\gamma^{\nu}D_{\mu}\psi$. Thus, matricial renormalization is required, namely
\begin{equation}
\mathcal{J}_0=\mathcal{Z}_{\mathcal{J}}\mathcal{J}\;,\label{j}
\end{equation}
where $\mathcal{J}$ is a column matrix of sources that share the same quantum numbers. The quantity $Z_{\mathcal{J}}$ is a squared matrix with the associated renormalization factors. In this case,
\begin{eqnarray}
\mathcal{J}_1=\begin{pmatrix}
\bar{\kappa}_{\alpha\mu\beta\nu}\\
C_{\nu\mu}
\end{pmatrix}\;&\mathrm{and}&\;\mathcal{Z}_1=\begin{pmatrix}
Z_{\bar{\kappa}\bar{\kappa}}& Z_{\bar{\kappa}C}\\
Z_{C\bar{\kappa}}& Z_{CC}
\end{pmatrix}\;\;=\;\;1+\varepsilon\mathbb{A}\;,
\label{renm1}
\end{eqnarray}
where $\mathbb{A}$ is a matrix depending on $a_i$. Thus
\begin{eqnarray}
\begin{pmatrix}
\bar{\kappa}_{0\alpha\mu\beta\nu}\\
C_{0\nu\mu}
\end{pmatrix}&=&\begin{pmatrix}
(Z_{\bar{\kappa}\bar{\kappa}})_{\alpha\mu\beta\nu}^{\phantom{\alpha\mu\beta\nu}\theta\rho\omega\delta}& (Z_{\bar{\kappa} C})_{\alpha\mu\beta\nu}^{\phantom{\alpha\mu\beta\nu}\theta\omega}\\
(Z_{C\bar{\kappa}})_{\nu\mu}^{\phantom{\nu\mu}\theta\rho\omega\delta}&(Z_{CC})_{\nu\mu}^{\phantom{\nu\mu}\theta\omega}
\end{pmatrix}\begin{pmatrix}
\bar{\kappa}_{\theta\rho\omega\delta}\\
C_{\theta\omega}
\end{pmatrix}\;,\nonumber\\
&=&\begin{pmatrix}
(1+\varepsilon(a_1-a_0))\delta^{\theta}_{\alpha}\delta^{\rho}_{\mu}\delta^{\omega}_{\beta}\delta^{\delta}_{\nu}& \varepsilon a_2 T_{\alpha\mu\beta\nu}^{\phantom{\alpha\mu\beta\nu}\theta\omega}\\
\varepsilon a_7\eta^{\rho\delta}\delta^{\theta}_{\nu}\delta^{\omega}_{\mu}&\delta^{\theta}_{\nu}\delta^{\omega}_{\mu}+\varepsilon((a_5-a_3)\delta^{\theta}_{\nu}\delta^{\omega}_{\mu}+a_6\delta^{\theta}_{\mu}\delta^{\omega}_{\nu})
\end{pmatrix}\begin{pmatrix}
\bar{\kappa}_{\theta\rho\omega\delta}\\
C_{\theta\omega}
\end{pmatrix}\;.\nonumber\\
\label{mr1}
\end{eqnarray}
As it is easy to infer from table \ref{table3}, some external sources do not have exactly the same quantum numbers, specifically with respect to their mass dimensions. Then, in principle, they do not suffer quantum mixing. However, the model has a mass parameter, the electron mass $m$. Thus, the mass parameter will enable extra mixing among sources \cite{Collins:1984xc}. Firstly,
\begin{eqnarray}
\begin{pmatrix}
\bar{A}^{\mu}_0\\
E^{\mu}_0
\end{pmatrix}&=&\begin{pmatrix}
Z_{\bar{A}\bar{A}}& Z_{\bar{A}E}\\
Z_{E\bar{A}}&Z_{EE}
\end{pmatrix}\begin{pmatrix}
\bar{A}^{\mu}\\
E^{\mu}
\end{pmatrix}\;,\nonumber\\
&=&\begin{pmatrix}
1+\varepsilon(a_{15}-a_3)&\varepsilon a_{16}m\\
0&1+\varepsilon(a_{10}-a_3)
\end{pmatrix}\begin{pmatrix}
\bar{A}^{\mu}\\
E^{\mu}
\end{pmatrix}\;.
\label{mr2}
\end{eqnarray}
Still
\begin{eqnarray}
\begin{pmatrix}
H_0^{\nu\mu}\\
D_0^{\nu\mu}
\end{pmatrix}&=&\begin{pmatrix}
(Z_{HH})^{\nu\mu\alpha\beta}& (Z_{HD})^{\nu\mu\alpha\beta}\\
(Z_{DH})^{\nu\mu\alpha\beta}&(Z_{DD})^{\nu\mu\alpha\beta}
\end{pmatrix}\begin{pmatrix}
H_{\alpha\beta}\\
D_{\alpha\beta}
\end{pmatrix}\;,\nonumber\\
&=&\begin{pmatrix}
(1+\varepsilon(a_{19}-a_3))\eta^{\nu\alpha}\eta^{\mu\beta}& \varepsilon a_{20}m\epsilon^{\nu\mu\alpha\beta}\\
0&\eta^{\nu\alpha}\eta^{\mu\beta}+\varepsilon((a_8-a_3)\eta^{\nu\alpha}\eta^{\mu\beta}+a_9\eta^{\nu\beta}\eta^{\mu\alpha})
\end{pmatrix}\begin{pmatrix}
H_{\alpha\beta}\\
D_{\alpha\beta}
\end{pmatrix}\;.\nonumber\\
\label{mr3}
\end{eqnarray}
The last renormalization factor is a mix among three sources, namely
\begin{eqnarray}
\begin{pmatrix}
B_0^{\mu}\\
J_0^{\alpha\beta\gamma}\\
G_0^{\alpha\beta\gamma}
\end{pmatrix}&=&\begin{pmatrix}
(Z_{BB})^{\mu}_{\phantom{\mu}\omega}& (Z_{BJ})^{\mu}_{\phantom{\mu}\lambda\rho\sigma}&(Z_{BG})^{\mu}_{\phantom{\mu}\lambda\rho\sigma}&\\
(Z_{JB})^{\alpha\beta\gamma}_{\phantom{\alpha\beta\gamma}\omega}&(Z_{JJ})^{\alpha\beta\gamma}_{\phantom{\alpha\beta\gamma}\lambda\rho\sigma}&(Z_{JG})^{\alpha\beta\gamma}_{\phantom{\alpha\beta\gamma}\lambda\rho\sigma}\\
(Z_{GB})^{\alpha\beta\gamma}_{\phantom{\alpha\beta\gamma}\omega}&(Z_{GJ})^{\alpha\beta\gamma}_{\phantom{\alpha\beta\gamma}\lambda\rho\sigma}&(Z_{GG})^{\alpha\beta\gamma}_{\phantom{\alpha\beta\gamma}\lambda\rho\sigma}
\end{pmatrix}\begin{pmatrix}
B^{\omega}\\
J^{\lambda\rho\sigma}\\
G^{\lambda\rho\sigma}
\end{pmatrix}\;,\nonumber\\
&=&\mathcal{Z}_4\begin{pmatrix}
B^{\mu}\\
J^{\alpha\beta\gamma}\\
G^{\alpha\beta\gamma}
\end{pmatrix}\;,
\label{mr4}
\end{eqnarray}
where
\begin{equation}
\mathcal{Z}_4=\begin{pmatrix}(1+\varepsilon(a_{17}-a_3))\delta^{\mu}_{\omega}&-\varepsilon a_{26} \epsilon_{\lambda\rho\sigma}^{\phantom{\lambda\rho\sigma}\mu}&\varepsilon a_{18}m\epsilon_{\lambda\rho\sigma}^{\phantom{\lambda\rho\sigma}\mu}\\
0&(1-\varepsilon a_0)\delta^{\alpha}_{\lambda}\delta^{\beta}_{\rho}\delta^{\gamma}_{\sigma}&0\\
0&0&\delta^{\alpha}_{\lambda}\delta^{\beta}_{\rho}\delta^{\gamma}_{\sigma}+\varepsilon((a_{12}-a_3)\delta^{\alpha}_{\lambda}\delta^{\beta}_{\rho}\delta^{\gamma}_{\sigma}+a_{13}S^{\alpha\beta\gamma}_{\phantom{\alpha\beta\gamma}\lambda\rho\sigma})
\end{pmatrix}\;.
\label{pm3}
\end{equation}
The renormalization for the external sources that do not suffer quantum mixing is the following
\begin{eqnarray}
Z_F&=&1+\varepsilon(a_{11}-a_3)\;,\nonumber\\
Z_{M_5}&=&1+\varepsilon(a_{14}-a_3)\;.
\end{eqnarray}
The bosonic sector associated with the $v^\mu$ vector, renormalizes through $Z_{JJ}$. It was already determined in \eqref{pm3}. Therefore, it has the following renormalization constraint
\begin{equation}
Z_{JJ}=Z^2_\lambda=Z_A^{-1}\;.
\label{ren3a}
\end{equation}

This ends the multiplicative renormalizability proof of the Lorentz-violating QED. We can see that, besides the three usual renormalizations of standard QED (related to $a_0$, $a_3$ and $a_4$), we also have nineteen extra parameters, associated to the breaking sector. Moreover, there are extra renormalizations associated with vacuum divergences (see Appendix \ref{Vacuumterms}). Thus, we achieve a total of twenty two independent renormalization parameters at dynamical sector of QED extension. We stress out that, as a consequence of the ghost Ward identity \eqref{19} (a feature of the Landau gauge), the term $\epsilon_{\mu\nu\alpha\beta}v^\mu A^\nu\partial^\alpha A^\beta$ does not renormalize. 

\section{One-loop computations}\label{ONELOOP}

\begin{figure}[htb]
	\centering
		\includegraphics[trim = 10mm 170mm 20mm 25mm, clip,width=0.60\textwidth]{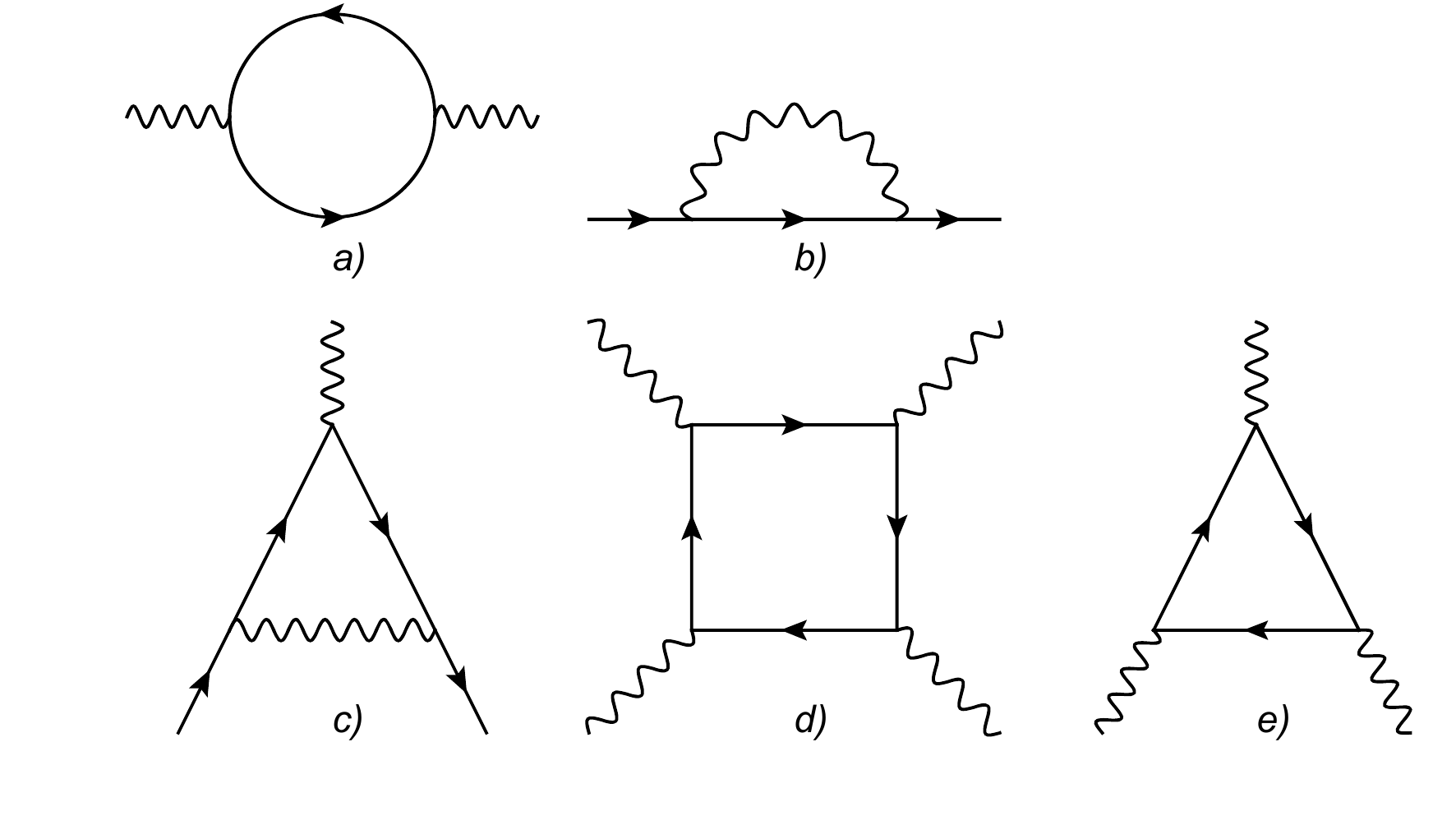}
		\caption{One-loop graphs for usual QED.}
		\label{fig1}
\end{figure}

As shown in the previous section, this model is renormalizable at least to all orders in perturbation theory. Even though the model presented here originates from underlying fundamental theories, the fact that the theory can be renormalized allows explicit consistent computations. In this section, we will study the QED extension diagrams, \emph{i.e.}, analyze the renormalizability in the diagrammatic scenario. In the context of Feynman diagrams, a quantum field theory is renormalizable whether divergences that arise in a one-particle irreducible (1PI) graph might be absorbed by redefinitions of the fields, parameters and coupling constants. Thus, we need to identify the superficial divergence degree presented in the QED extension. From Feynman rules of this model (see appendix \ref{FR}) \cite{Kostelecky:2001jc}, the degree of divergence is given by
\begin{eqnarray}
D=4-B-\frac{3}{2}F-V_{B}-V_{F}\ ,
\label{37.A}
\end{eqnarray}
where $B$ is the number of bosonic external legs, $F$ the number of fermionic external legs, $V_{B}$ is the massive insertion at the bosonic propagator and $V_{F}$ is the massive insertion at the fermionic propagator. The usual QED presents a finite number of divergent diagrams, see Fig.~\ref{fig1}. However, through Ward identities, it is possible to show that the $d)$ graph does not present any divergence - This can be directly seen from the counterterm \eqref{31}. And, by Furry's theorem \cite{Furry:1951zz}, the $e)$ graph has a total vanishing amplitude.

\begin{figure}[htb]
	\centering
		\includegraphics[trim = 20mm 195mm 20mm 25mm, clip,width=0.60\textwidth]{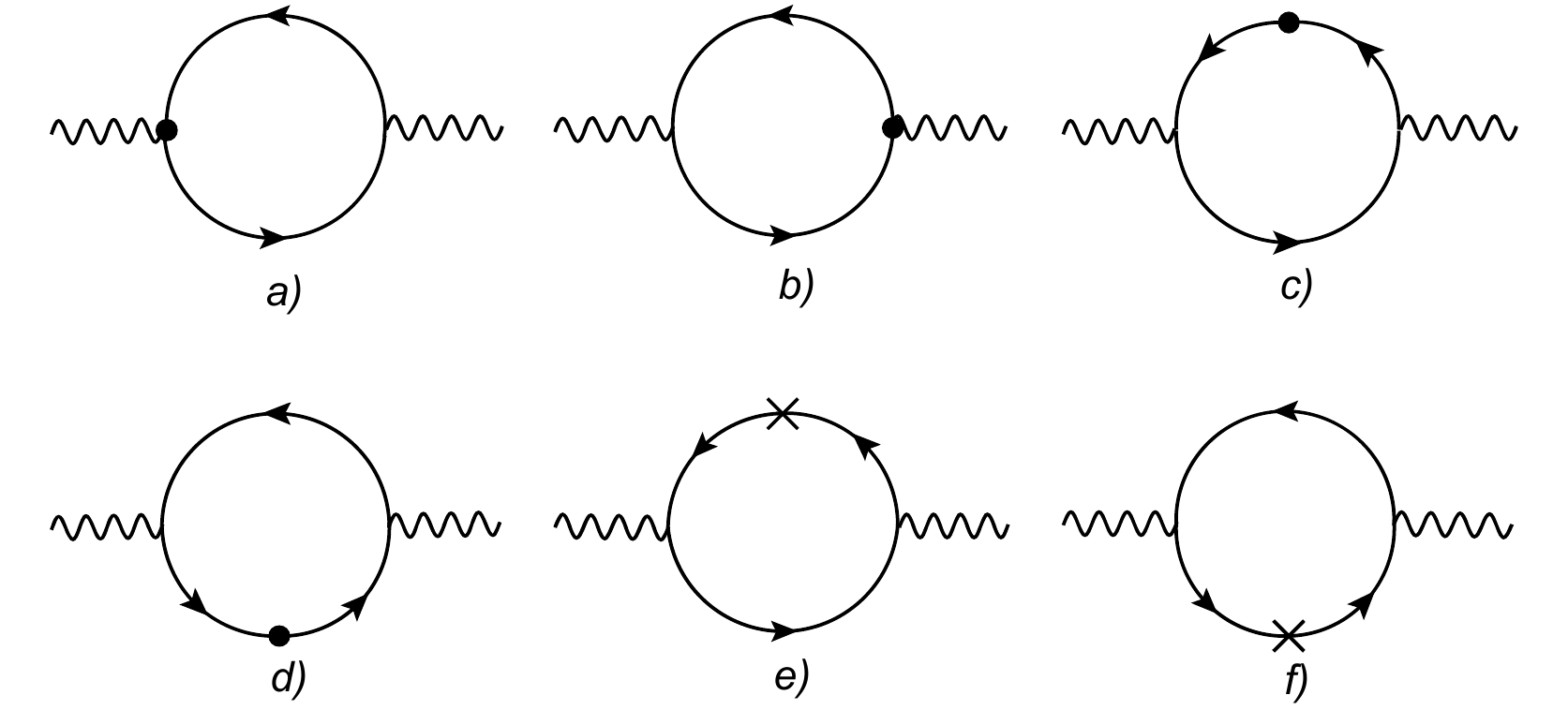}
		\caption{One-loop vacuum polarization in the QED extension.}
		\label{fig2}
\end{figure}

\begin{figure}[htb]
	\centering
		\includegraphics[trim = 20mm 210mm 20mm 25mm, clip,width=0.60\textwidth]{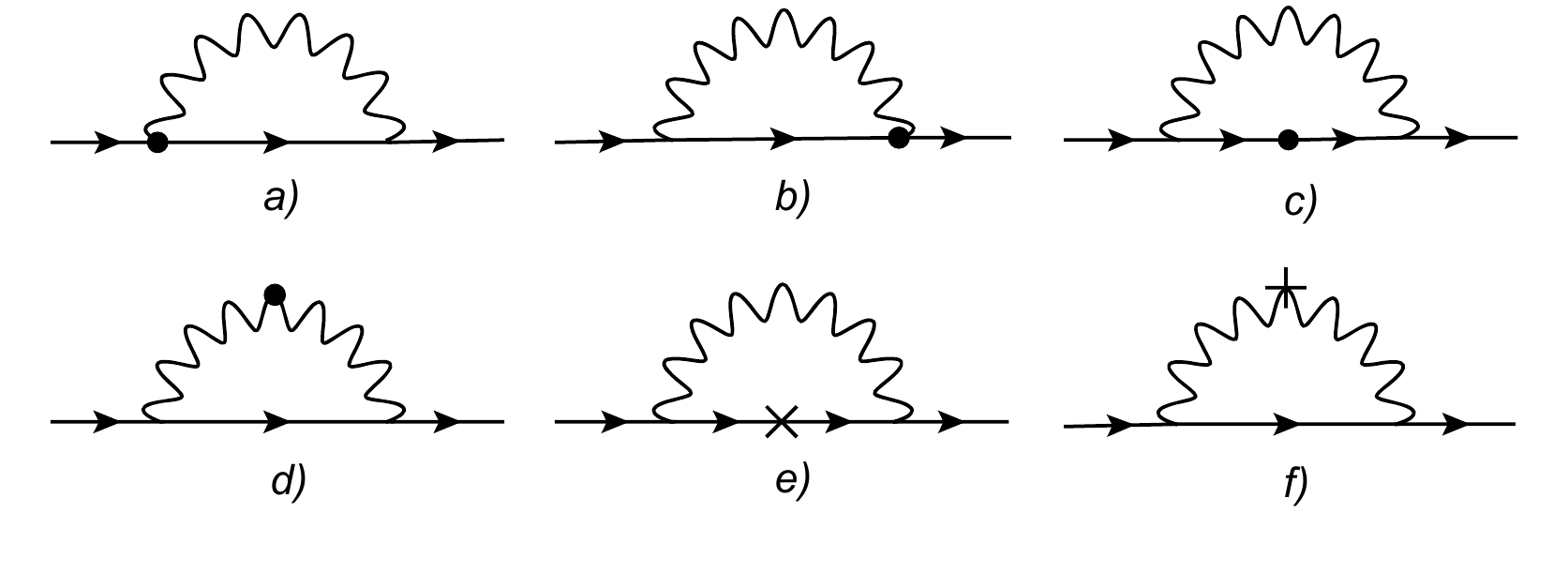}
		\caption{One-loop self-energy of the electron in the QED extension.}
		\label{fig3}
\end{figure}

\begin{figure}[htb]
	\centering
		\includegraphics[trim = 20mm 45mm 20mm 25mm, clip,width=0.40\textwidth]{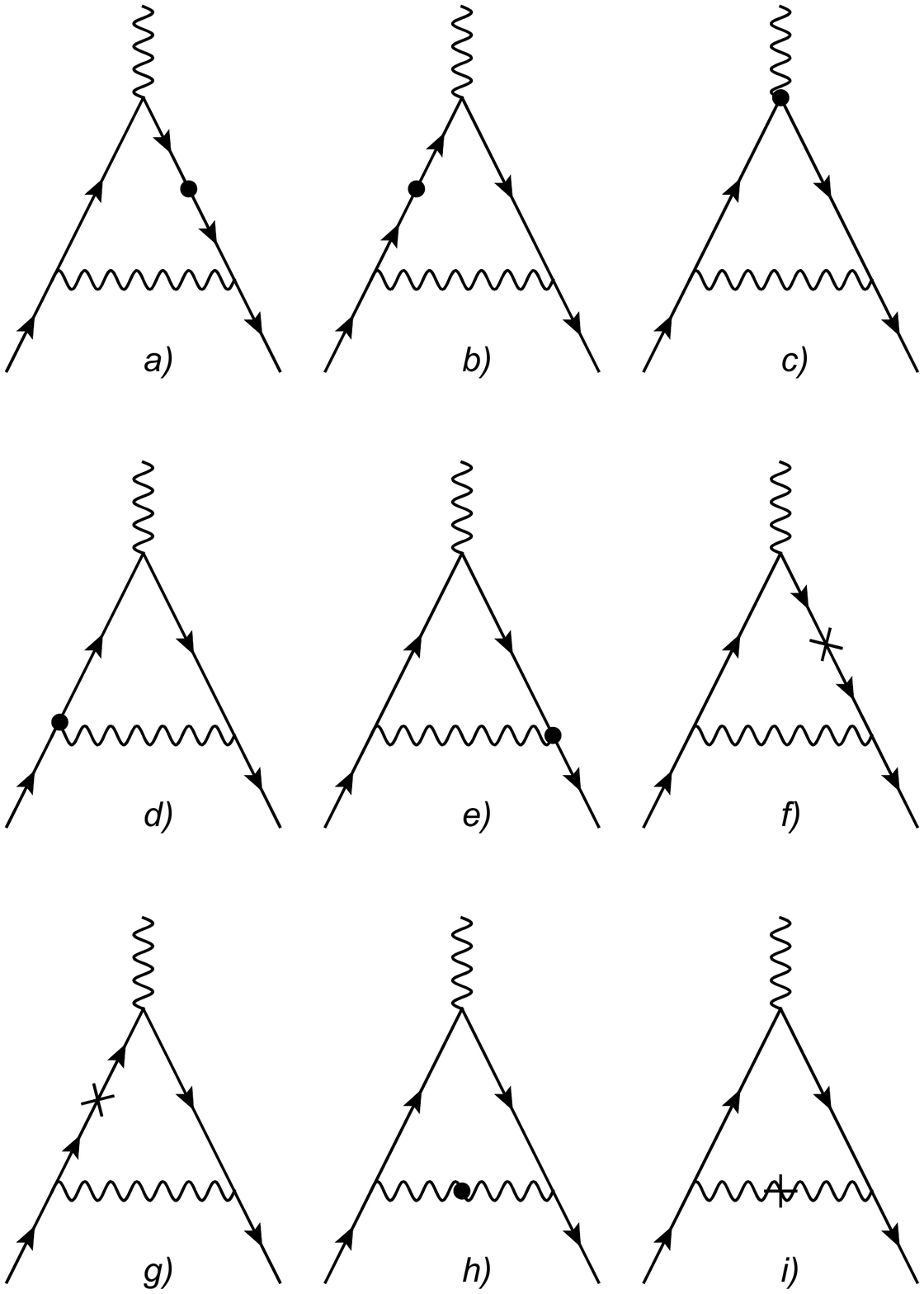}
		\caption{Fermion-photon vertices in the QED extension.}
		\label{fig4}
\end{figure}

With respect to the Lorentz-violating diagrams, they can be obtained by single introduction of the Lorentz violation coefficients (on the potentially divergent graphs presented) as insertions in the usual QED \cite{Kostelecky:2001jc}. Such topologically inequivalent diagrams are shown in Figs.~\ref{fig2}, \ref{fig3} and \ref{fig4}. With respect to the insertions in the three-photon vertex, they are outside of the scope of this work\footnote{The reader can see for instance \cite{Franco:2013rp}, where the authors claim the absence of gauge anomalies at one-loop order, and point out the possibility of the renormalizability of the Lorentz-violating QED to all orders.}. 
In order to compute the graphs that present divergences in high momenta, we need to employ a method to regularize this graphs. Here, we adopt dimensional regularization \cite{Itzykson:1980rh,Bollini:1972bi,'tHooft:1972fi}, by performing Feynman integrals computations in arbitrary dimensions $d=4-\epsilon$, with poles at $\epsilon=0$. Even though the model breaks Lorentz invariance, dimensional regularization is quite useful because it does not refer to Poincar\'e invariance. Furthermore, as we already mentioned before, the free propagators are Lorentz invariants, while the breaking coefficients are just insertions. 

The radiative corrections to the graphs shown in Fig.~\ref{fig1} for vacuum polarization, electron self-energy and fermion-photon vertex, at one-loop, are respectively given by
\begin{eqnarray}
\Pi^{\mu\nu}_{a}(p)&=&\frac{4}{3}I_{0}\left(p^{\mu}p^{\nu}-p^{2}\eta^{\mu\nu}\right)\;,\nonumber\\
\Sigma_{b}(p)&=&I_{0}\left(-\gamma^{\mu}p_{\mu}+4m\right)\;,\nonumber\\
\Lambda^{\mu}_{c}&=&I_{0}\gamma^{\mu}\;,
\end{eqnarray}
where $I_{0}=\frac{e^2}{8\pi^2\epsilon}$.
For the graphs of the vacuum polarization shown in the Fig.~\ref{fig2}, one finds
\begin{align}
\Pi^{\mu\nu}_{a}(p)&=\frac{4}{3}I_{0}\left(c^{\alpha\mu}p_{\alpha}p^{\nu}-p^{2}c^{\nu\mu}-6mig^{\alpha\nu\mu}p_{\alpha}\right)\;,\displaybreak[3]\nonumber\\
\Pi^{\mu\nu}_{b}(p)&=\frac{4}{3}I_{0}\left(c^{\alpha\nu}p_{\alpha}p^{\mu}-p^{2}c^{\mu\nu}-6mig^{\mu\alpha\nu}p_{\alpha}\right)\;,\displaybreak[3]\nonumber\\
\Pi^{\mu\nu}_{c}(p)&=\frac{2}{3}I_{0}\left[c^{\mu\alpha}p_{\alpha}p^{\nu}+c^{\nu\alpha}p_{\alpha}p^{\mu}-2c^{\alpha\beta}p_{\alpha}p_{\beta}\eta^{\mu\nu}+m\left(p^{\mu}e^{\nu}+p^{\nu}e^{\mu}-2e^{\alpha}p_{\alpha}\eta^{\mu\nu}\right)+\right.\nonumber\\
&+\left.6mi\left(g^{\alpha\nu\mu}+g^{\mu\alpha\nu}\right)p_{\alpha}\right]\;,\displaybreak[3]\nonumber\\
\Pi^{\mu\nu}_{d}(p)&=\frac{2}{3}I_{0}\left[c^{\mu\alpha}p_{\alpha}p^{\nu}+c^{\nu\alpha}p_{\alpha}p^{\mu}-2c^{\alpha\beta}p_{\alpha}p_{\beta}\eta^{\mu\nu}-m\left(p^{\mu}e^{\nu}+p^{\nu}e^{\mu}-2e^{\alpha}p_{\alpha}\eta^{\mu\nu}\right)+\right.\nonumber\\
&+\left.6mi\left(g^{\alpha\nu\mu}+g^{\mu\alpha\nu}\right)p_{\alpha}\right]\;,\displaybreak[3]\nonumber\\
\Pi^{\mu\nu}_{e}(p)&=\frac{4}{3}I_{0}\left(p^{\mu}a^{\nu}+p^{\nu}a^{\mu}-2a^{\alpha}p_{\alpha}\eta^{\mu\nu}+6mih^{\mu\nu}\right)\;,\displaybreak[3]\nonumber\\
\Pi^{\mu\nu}_{f}(p)&=-\frac{4}{3}I_{0}\left(p^{\mu}a^{\nu}+p^{\nu}a^{\mu}-2a^{\alpha}p_{\alpha}\eta^{\mu\nu}+6mih^{\mu\nu}\right)\;.
\label{FP}
\end{align}
Although each graph of the vacuum polarization receives contributions from anti-symmetric pieces of the Lorentz-violating coefficients of the fermion sector, and also, massive contributions from the fermion mass, the overall contribution, as shown in \cite{Kostelecky:2001jc}, is symmetric and independent of the fermion mass. 

The corrections related to the self-energy of the electron in the QED extension (Fig.~\ref{fig3}) are given by
\begin{align}
\Sigma_a(p)&=I_0\left[\left(\Gamma^{\mu}-\frac{1}{2}\Gamma^{\nu}\gamma_{\nu}\gamma^{\mu}\right)p_{\mu}+m\Gamma^{\nu}\gamma_{\nu}\right]\;,\displaybreak[3]\nonumber\\
\Sigma_b(p)&=I_0\left[\left(c^{\mu\nu}+d^{\mu\nu}\gamma_{5}\right)\gamma_{\nu}p_{\mu}-g^{\mu\alpha\beta}\sigma_{\alpha\beta}p_{\mu}+g^{\beta}_{\ \alpha\beta}\sigma^{\alpha\mu}p_{\mu}+\frac{1}{2}\Gamma^{\nu}\gamma_{\nu}\gamma^{\mu}p_{\mu}+\right.\nonumber\\
&-\left.m\left(c^{\nu\mu}\gamma_{\nu}\gamma_{\mu}-d^{\nu\mu}\gamma_{5}\gamma_{\nu}\gamma_{\mu}-e^{\mu}\gamma_{\mu}+if^{\mu}\gamma_{\mu}-\frac{1}{2}g^{\alpha\beta\mu}\gamma_{\mu}\sigma_{\alpha\beta}\right)\right]\;,\displaybreak[3]\nonumber\\
\Sigma_c(p)&=I_0\left[\frac{1}{3}\left(c^{\mu\nu}+d^{\mu\nu}\gamma_{5}\right)\gamma_{\nu}p_{\mu}-\frac{2}{3}\left(c^{\nu\mu}+d^{\nu\mu}\gamma_{5}\right)\gamma_{\nu}p_{\mu}-2\left(e^{\mu}+if^{\mu}\gamma_{5}\right)p_{\mu}+\right.\nonumber\\
&+\left.m\left(e^{\mu}\gamma_{\mu}-\frac{1}{4}g^{\alpha\beta\mu}\gamma_{\mu}\sigma_{\alpha\beta}-\frac{1}{4}g^{\alpha\beta\mu}\sigma_{\alpha\beta}\gamma_{\mu}\right)\right]\;,\displaybreak[3]\nonumber\\
\Sigma_d(p)&=I_0\left[4im_{5}\gamma_{5}+\left(a^{\mu}+b^{\mu}\gamma_{5}\right)\gamma_{\mu}\right],\displaybreak[3]\nonumber\\
\Sigma_e(p)&=3I_0\gamma_{5}\gamma_{\mu}v^{\mu}\;,\displaybreak[3]\nonumber\\
\Sigma_f(p)&=-\frac{4}{3}I_0\gamma^{\nu}p_{\mu}\kappa^{\mu\alpha}_{\phantom{\mu\alpha}\nu\alpha}\;.
\label{SE}
\end{align}
The overall contribution is consistent with that shown in \cite{Kostelecky:2001jc}, except for the additional piece shown at the correction to the $d)$ graph. This additional piece comes from the first massive Lorentz and CPT violation insertion at \eqref{3}, namely, the Lorentz-violating coefficient coupled to the pseudo-scalar current, $m_5$. This term was not present in the computation presented at \cite{Kostelecky:2001jc}. Indeed, this term was added to Lorentz-violating action at \cite{DelCima:2012gb}.

Finally, the one-loop corrections for the electron-photon vertex in the QED extension (Fig.~\ref{fig4}) are
\begin{align}
\Lambda^{\mu}_a&=-I_0\left[\frac{2}{3}\left(c^{\nu\mu}+d^{\nu\mu}\gamma_5\right)\gamma_{\nu}-\frac{1}{2}\left(c^{\nu\alpha}+d^{\nu\alpha}\gamma_5\right)\gamma_{\nu}\gamma_{\alpha}\gamma^{\mu}-\frac{1}{3}\left(c^{\mu\nu}+d^{\mu\nu}\gamma_5\right)\gamma_{\nu}+e^{\mu}+\right.\nonumber\\
&+\left.if^{\mu}\gamma_{5}+\frac{1}{2}g^{\alpha\beta\mu}\sigma_{\alpha\beta}-\frac{1}{4}g^{\alpha\beta\sigma}\left(\sigma_{\alpha\beta}\gamma_{\sigma}\gamma^{\mu}+\gamma_{\sigma}\gamma^{\mu}\sigma_{\alpha\beta}\right)\right]\;, \displaybreak[3] \nonumber\\
\Lambda^{\mu}_b&=-I_0\left[\frac{2}{3}\left(c^{\mu\nu}+d^{\mu\nu}\gamma_5\right)\gamma_{\nu}+\frac{1}{2}\left(c^{\nu\alpha}+d^{\nu\alpha}\gamma_5\right)\gamma_{\nu}\gamma_{\alpha}\gamma^{\mu}-\frac{1}{3}\left(c^{\nu\mu}+d^{\nu\mu}\gamma_5\right)\gamma_{\nu}+e^{\mu}+\right.\nonumber\\
&+\left.if^{\mu}\gamma_{5}-\frac{1}{2}g^{\alpha\beta\mu}\sigma_{\alpha\beta}+\frac{1}{4}g^{\alpha\beta\sigma}\left(\sigma_{\alpha\beta}\gamma_{\sigma}\gamma^{\mu}+\gamma_{\sigma}\gamma^{\mu}\sigma_{\alpha\beta}\right)\right]\;, \displaybreak[3] \nonumber\\ 
\Lambda^{\mu}_c&=I_0\left[\left(c^{\nu\mu}+d^{\nu\mu}\gamma_5\right)\gamma_{\nu}+4\left(e^{\mu}+if^{\mu}\gamma_{5}\right)\right]\;,\displaybreak[3]\nonumber\\
\Lambda^{\mu}_d&=-I_0\left(\Gamma^{\mu}-\frac{1}{2}\Gamma^{\nu}\gamma_{\nu}\gamma^{\mu}\right)\;, \displaybreak[3]\nonumber\\
\Lambda^{\mu}_e&=-I_0\left[\left(c^{\mu\nu}+d^{\mu\nu}\gamma_5\right)\gamma_{\nu}+\frac{1}{2}\left(c^{\nu\alpha}+d^{\nu\alpha}\gamma_5\right)\gamma_{\nu}\gamma_{\alpha}\gamma^{\mu}+\frac{1}{2}\left(e^{\nu}+if^{\nu}\gamma_5\right)\gamma_{\nu}\gamma^{\mu}+\right.\nonumber\\
&+\left.\frac{1}{4}g^{\alpha\beta\sigma}\gamma_{\sigma}\gamma^{\mu}\sigma_{\alpha\beta}\right]\;, \displaybreak[3]\nonumber\\
\Lambda^{\mu}_i&=\frac{4}{3}I_0\gamma^{\nu}\kappa^{\mu\alpha}_{\phantom{\mu\alpha}\nu\alpha}\;, \displaybreak[3]\nonumber\\
\Lambda^{\mu}_f&=\Lambda^{\mu}_g=\Lambda^{\mu}_h=0\;.\displaybreak[3]
\label{EPV}
\end{align}
Again, the overall contribution for the electron-photon vertex in the QED extension is consistent with that shown in \cite{Kostelecky:2001jc}. 

In order to compare the explicit one-loop results with the algebraic results in the previous section, we can now compute the explicit renormalizations factors at one-loop order. For the renormalization factors for the photon, electron and electron mass, it is found
\begin{eqnarray}
Z_A^{1/2}&=&1-\frac{e^2}{12\pi^2\epsilon}\;,\nonumber\\
Z_\psi^{1/2}&=&1-\frac{e^2}{16\pi^2\epsilon}\;,\nonumber\\
Z_m&=&1-\frac{3e^2}{8\pi^2\epsilon}\;.
\label{ren1A}
\end{eqnarray}
The renormalization for electric charge is given by
\begin{equation}
Z_e=1+\frac{e^2}{12\pi^2\epsilon}=Z_{A}^{-1/2}\;.
\label{ren1B}
\end{equation}

For the local sources, taking their physical values, it is also needed to employ matrix renormalization. Firstly, for the renormalization matrix of the tensors $\kappa_{\alpha\mu\beta\nu}$ and $c_{\nu\mu}$, like in \eqref{mr1}, we obtain
\begin{eqnarray}
\begin{pmatrix}
\kappa_{0\alpha\mu\beta\nu}\\
c_{0\nu\mu}
\end{pmatrix}&=&\begin{pmatrix}
(Z_{\kappa\kappa})_{\alpha\mu\beta\nu}^{\phantom{\alpha\mu\beta\nu}\theta\rho\omega\delta}& (Z_{\kappa c})_{\alpha\mu\beta\nu}^{\phantom{\alpha\mu\beta\nu}\theta\omega}\\
(Z_{c\kappa})_{\nu\mu}^{\phantom{\nu\mu}\theta\rho\omega\delta}&(Z_{cc})_{\nu\mu}^{\phantom{\nu\mu}\theta\omega}
\end{pmatrix}\begin{pmatrix}
\kappa_{\theta\rho\omega\delta}\\
c_{\theta\omega}
\end{pmatrix}\;,\nonumber\\
&=&\begin{pmatrix}
(1+\frac{e^2}{6\pi^2\epsilon})\delta^{\theta}_{\alpha}\delta^{\rho}_{\mu}\delta^{\omega}_{\beta}\delta^{\delta}_{\nu}& \frac{e^2}{12\pi^2\epsilon}T_{\alpha\mu\beta\nu}^{\phantom{\alpha\mu\beta\nu}\theta\omega}\\
-\frac{e^2}{6\pi^2\epsilon}\eta^{\rho\delta}\delta^{\theta}_{\nu}\delta^{\omega}_{\mu}&\delta^{\theta}_{\nu}\delta^{\omega}_{\mu}+\frac{e^2}{6\pi^2\epsilon}(\delta^{\theta}_{\nu}\delta^{\omega}_{\mu}+\delta^{\theta}_{\mu}\delta^{\omega}_{\nu})
\end{pmatrix}\begin{pmatrix}
\kappa_{\theta\rho\omega\delta}\\
c_{\theta\omega}
\end{pmatrix}\;,
\end{eqnarray}
Just like shown in \eqref{mr2}, for $a^{\mu}$ and $e^{\mu}$ tensors, it is obtained
\begin{eqnarray}
\begin{pmatrix}
a^{\mu}_0\\
e^{\mu}_0
\end{pmatrix}&=&\begin{pmatrix}
Z_{aa}& Z_{ae}\\
Z_{ea}&Z_{ee}
\end{pmatrix}\begin{pmatrix}
a^{\mu}\\
e^{\mu}
\end{pmatrix}\;,\nonumber\\
&=&\begin{pmatrix}
1& -\frac{3e^2}{8\pi^2\epsilon}m\\
0&1
\end{pmatrix}\begin{pmatrix}
a^{\mu}\\
e^{\mu}
\end{pmatrix}\;,
\end{eqnarray}
For $h^{\nu\mu}$ and $d^{\nu\mu}$ tensors, just like in \eqref{mr3}, we have
\begin{eqnarray}
\begin{pmatrix}
h_0^{\nu\mu}\\
d_0^{\nu\mu}
\end{pmatrix}&=&\begin{pmatrix}
(Z_{hh})^{\nu\mu\alpha\beta}& (Z_{hd})^{\nu\mu\alpha\beta}\\
(Z_{dh})^{\nu\mu\alpha\beta}&(Z_{dd})^{\nu\mu\alpha\beta}
\end{pmatrix}\begin{pmatrix}
h_{\alpha\beta}\\
d_{\alpha\beta}
\end{pmatrix}\;,\nonumber\\
&=&\begin{pmatrix}
(1+\frac{e^2}{8\pi^2\epsilon})\eta^{\nu\alpha}\eta^{\mu\beta}& -m\frac{e^2}{4\pi^2\epsilon}\epsilon^{\nu\mu\alpha\beta}\\
0&\eta^{\nu\alpha}\eta^{\mu\beta}+\frac{e^2}{6\pi^2\epsilon}(\eta^{\nu\alpha}\eta^{\mu\beta}+\eta^{\nu\beta}\eta^{\mu\alpha})
\end{pmatrix}\begin{pmatrix}
h_{\alpha\beta}\\
d_{\alpha\beta}
\end{pmatrix}\;.
\end{eqnarray}
For $b^{\mu}$, $v^{\mu}$ and $g^{\alpha\beta\gamma}$ tensors, in accordance with \eqref{mr4}, we have the following renormalization matrix
\begin{eqnarray}
\begin{pmatrix}
b_0^{\mu}\\
v_0^{\mu}\\
g_0^{\alpha\beta\gamma}
\end{pmatrix}&=&\begin{pmatrix}
(Z_{bb})^{\mu}_{\phantom{\mu}\omega}& (Z_{bv})^{\mu}_{\phantom{\mu}\omega}&(Z_{bg})^{\mu}_{\phantom{\mu}\lambda\rho\sigma}&\\
(Z_{vb})^{\mu}_{\phantom{\mu}\omega}&(Z_{vv})^{\mu}_{\phantom{\mu}\omega}&(Z_{vg})^{\mu}_{\phantom{\mu}\lambda\rho\sigma}\\
(Z_{gb})^{\alpha\beta\gamma}_{\phantom{\alpha\beta\gamma}\omega}&(Z_{gv})^{\alpha\beta\gamma}_{\phantom{\alpha\beta\gamma}\omega}&(Z_{gg})^{\alpha\beta\gamma}_{\phantom{\alpha\beta\gamma}\lambda\rho\sigma}
\end{pmatrix}\begin{pmatrix}
b^{\omega}\\
v^{\omega}\\
g^{\lambda\rho\sigma}
\end{pmatrix}\;,\nonumber\\
&=&\begin{pmatrix}\delta^{\mu}_{\omega}&-\frac{3e^2}{8\pi^2\epsilon}\delta^{\mu}_{\omega}&m\frac{e^2}{16\pi^2\epsilon}\epsilon_{\lambda\rho\sigma}^{\phantom{\lambda\rho\sigma}\mu}\\
0&(1+\frac{e^2}{6\pi^2\epsilon})\delta^{\mu}_{\omega}&0\\
0&0&\delta^{\alpha}_{\lambda}\delta^{\beta}_{\rho}\delta^{\gamma}_{\sigma}+\frac{e^2}{8\pi^2\epsilon}(2\delta^{\alpha}_{\lambda}\delta^{\beta}_{\rho}\delta^{\gamma}_{\sigma}+S^{\alpha\beta\gamma}_{\phantom{\alpha\beta\gamma}\lambda\rho\sigma})
\end{pmatrix}\begin{pmatrix}
b^{\omega}\\
v^{\omega}\\
g^{\lambda\rho\sigma}
\end{pmatrix}\;.
\end{eqnarray}
Remarkably, the renormalization factor of the background $v^\mu$ is in accordance with \eqref{ren3a}. For the tensors that do not suffer quantum mixing, $f^{\mu}$ and $m_5$, we have the following renormalization factors
\begin{eqnarray}
Z_f&=&1\;,\nonumber\\
Z_{m_5}&=&1-\frac{3e^2}{8\pi^2\epsilon}\;.
\end{eqnarray}

A few comments are in order: In the previous section was shown the all orders renormalizability of Lorentz- and CPT-violating QED. From \eqref{ren2}, the equivalence between the photon and electric charge renormalizations, \textit{i.e.}, $Z_e=Z_A^{-1/2}$, is ensured. This is confirmed from explicit computation at one-loop order, see \eqref{ren1B}. Another interesting point is the transversality of the photon propagator. In fact, one-loop computations displayed at \eqref{FP} show this property. From the first Ward identity at \eqref{17}, this feature is ensured to all orders in perturbation theory. However, we were not able to fix, from the algebraic approach, a relation between the renormalization factors of the sources (background fields) and the renormalization factors of the photon and electron fields. In fact, from explicit computation, it is possible to see that, for instance, no renormalization is need for $f^{\mu}$, at least at one-loop order. This is intimately related to the renormalization of the electron field. Moreover, the two approaches -- algebraic and analytical -- show to us that quantum mixing between the background fields is unavoidable. Since the algebraic approach has not given us all restrictions on renormalization parameters of the sources which appears at the analytical relations, it still remain to establish if this is just a one-loop effect that disappears at higher order or there are extra symmetries not considered in our set of Ward identities.

Furthermore, we can take the action \eqref{31} at the physical limit of the sources \eqref{12} and explore some special cases. For instance, considering only the terms of the action \eqref{31} which remain invariant under PT symmetry, \textit{i.e.}, discarding the terms depending on $h^{\mu\nu}, d^{\mu\nu}, b^{\mu}, g^{\alpha\beta\mu}$ and $v^{\mu}$, we get
\begin{eqnarray}
\Sigma^c_{PT-inv}&=&-\frac{1}{4}\int d^4x\;a_0F^{\mu\nu}F_{\mu\nu}-\frac{1}{4}\int d^4x\left(a_1\kappa_{\alpha\mu\beta\nu}+a_2T_{\alpha\mu\beta\nu}^{\phantom{\alpha\mu\beta\nu}\theta\omega}c_{\theta\omega}\right)F^{\alpha\mu}F^{\beta\nu}+\nonumber\\
&+&\int d^4x\left\{a_3i\overline{\psi}\gamma^{\mu}D_{\mu}\psi-a_4m\overline{\psi}\psi+ i\left[(a_5c^{\nu\mu}+ a_6c^{\mu\nu}+a_7\eta_{\alpha\beta}\kappa^{\alpha\mu\beta\nu})\overline{\psi}\gamma_{\nu}D_{\mu}\psi+\right.\right.\nonumber\\
&+&\left.\left.a_{10}e^{\mu}\overline{\psi}D_{\mu}\psi+a_{11}if^{\mu}\overline{\psi}\gamma_{5}D_{\mu}\psi\right]-\left[a_{14}im_5\overline{\psi}\gamma_5\psi+\left(a_{15}a^{\mu}+a_{16} me^{\mu}\right)\overline{\psi}\gamma_{\mu}\psi\right]\right\}\;.\nonumber\\
\label{PTI}
\end{eqnarray}
It is not difficult to see that this action remains stable, with thirteen renormalization parameters. On the other hand, if we choose C-invariance for the counterterm, we get
\begin{eqnarray}
\Sigma^c_{C-inv}&=&-\frac{1}{4}\int d^4x\;a_0F^{\mu\nu}F_{\mu\nu}-\frac{1}{4}\int d^4x\left(a_1\kappa_{\alpha\mu\beta\nu}+a_2T_{\alpha\mu\beta\nu}^{\phantom{\alpha\mu\beta\nu}\theta\omega}c_{\theta\omega}\right)F^{\alpha\mu}F^{\beta\nu}+\nonumber\\
&+&\int d^4x\left\{a_3i\overline{\psi}\gamma^{\mu}D_{\mu}\psi-a_4m\overline{\psi}\psi+ i\left[(a_5c^{\nu\mu}+ a_6c^{\mu\nu}+a_7\eta_{\alpha\beta}\kappa^{\alpha\mu\beta\nu})\overline{\psi}\gamma_{\nu}D_{\mu}\psi+\right.\right.\nonumber\\
&+&\left.\left.\frac{1}{2}\left(a_{12}g^{\alpha\beta\gamma }+a_{13}S^{\alpha\beta\gamma}_{\phantom{\alpha\beta\gamma}\lambda\rho\sigma}g^{\lambda\rho\sigma}\right)\overline{\psi}\sigma_{\alpha\beta}D_{\gamma}\psi\right]-\left[a_{14}im_5\overline{\psi}\gamma_5\psi+\right.\right.\nonumber\\
&+&\left.\left.(a_{17}b^{\mu}+a_{18}mg_{\alpha\beta\gamma}\epsilon^{\alpha\beta\gamma\mu}-6a_{26}v^{\mu})\overline{\psi}\gamma_5\gamma_{\mu}\psi\right]\right\}\;.
\label{CI}
\end{eqnarray}
Which is also stable under matrix renormalization with fourteen independent parameters.

\section{Conclusion}\label{FINAL}

In this work we have shown the multiplicative renormalizability of the general Lorentz- and CPT-violating quantum electrodynamics, at least to all orders in perturbation theory. We employed the algebraic renormalization technique in the BRST formalism in combination with a set of external sources which controls the breaking terms. In fact, a detailed study of the solutions of the Slavnov-Taylor operator in the space of local integrated polynomial in the fields of ghost number one and dimension bounded by four has been performed, and the Lorentz-violating QED is expected to be free of gauge anomalies to all orders in perturbation theory \cite{Tiago}. With this approach we have found the following results:

\begin{itemize}

\item  The renormalizability features of the standard QED sector is left unchanged, see \eqref{ren1} and \eqref{ren2}.

\item We have found a total amount of twenty two parameters for the action which respects Lorentz, CPT and BRST symmetries. When we set the physical values of the sources \eqref{12} and, for instance, choose PT-invariance, we found thirteen parameters, instead of the nine parameters found in \cite{DelCima:2012gb}. The discrepancy between these results is due to the matrix renormalization employed. We stress out that matrix renormalization is not a choice, but a need due to the mixing among sources. These results are consistent with the one-loop computations developed in Section \ref{ONELOOP}. In fact, the one-loop computations here developed generalize those presented in \cite{Kostelecky:2001jc} by including all terms considered in \cite{DelCima:2012gb} and a few more.

\item The Carroll-Field-Jackiw term $\epsilon_{\mu\nu\alpha\beta}v^{\mu}A^{\nu}\partial^{\alpha}A^{\beta}$ does not renormalize. Moreover, this is confirmed by one-loop explicit computations. In fact, this is direct a consequence of the ghost Ward identity \eqref{19}. 

\item In contrast to the non-Abelian case \cite{Santos:2014lfa}, Proca-like terms are not generated by the Lorentz-violating coefficients. This is also a consequence of the ghost equation which, at the Abelian case, is not an integrated identity, making it stronger than its non-Abelian version.

\item Although the Lorentz-violating coefficients are phenomenologically tiny, we found from our approach that the vacuum of the model is modified. In fact, vacuum terms are not avoided by the Ward identities. Anyhow, since these terms are pure source terms, the dynamical content of the model is maintained. Thus, attributes as causality and unitarity are also preserved \cite{Adam:2001ma}. Nevertheless, it is worth mentioning a few words about the vacuum terms. Typically, vacuum terms are associated to condensates, from spontaneous symmetry breaking mechanisms \cite{Higgs:1964ia,Guralnik:1964eu}, dynamical condensation effects \cite{Dudal:2011gd,Fukuda:1977mz,Gusynin:1978tr,Cornwall:1981zr,Verschelde:2001ia}, and so on. In here, a slightly different effect takes place, which is similar to the Gribov-Zwanziger vacuum terms \cite{Zwanziger:1992qr,Dudal:2005na,Maggiore:1993wq}, \textit{i.e.}, these terms come from the pure source terms allowed by power counting and survive after the physical limit. We can interpret these terms as follows: Because the theory is embeded in a larger theory, the contraction to the physical theory leaves a ``memory'' of the larger action. It can however be understood as a condensation of the classical set of fields when the physical limit is taken, \textit{i.e.}, the physical limit is a non-trivial freezing of these auxiliary fields. We can also remark that the Gribov-Zwanziger action can be interpreted as a spontaneous symmetry broken action \cite{Dudal:2012sb}, perhaps a similar interpretation is possible for the present approach.

\item Finally, we have found that there is no radiative generation of the Chern-Simons-like term $\epsilon_{\mu\nu\alpha\beta}b^{\mu}A^{\nu}\partial^{\alpha}A^{\beta}$ \cite{Jackiw:1999yp,Bonneau:2000ai,PerezVictoria:1999uh}. We have shown that this property is a direct consequence of the BRST classes of the sources employed in our approach: In order to control the Lorentz and CPT breaking, the background field $b^\mu$ was promoted to the external source $B^\mu$. Since this source is coupled to a BRST invariant composite operator, it is BRST closed. On the other hand, the background field $v^{\mu}$ was promoted to the external source $J^{\mu\nu\alpha}$. This source, however, is coupled to a BRST non-invariant operator. Hence, it must be BRST exact, with $\lambda^{\mu\nu\alpha}$ being its BRST counterpart in a BRST doublet. As BRST exact sources/operators can not receive contribution from BRST closed sources/operators \cite{Joglekar:1975nu,Henneaux:1993jn,Collins:1994ee}, the $J^{\mu\nu\alpha}$ source will never receive contribution from the $B^\mu$ source -- on the other hand, the other way is possible. See, for instance, the C-invariant counterterm \eqref{CI}, where the term $\epsilon_{\mu\nu\alpha\beta}b^{\mu}A^{\nu}\partial^{\alpha}A^{\beta}$ does not appear. Moreover, since the algebraic technique is a recursive and regularization scheme independent method, this property is ensured to all orders in perturbation theory. This result was also found at \cite{DelCima:2009ta}. However, here we found a cohomological interpretation.

\end{itemize}

\appendix

\section{Vacuum terms}\label{Vacuumterms}

We will discuss now the vacuum action, \textit{i.e.}, the action that taken account only terms depending on the sources. Since this action does not interfere with the renormalization of the sources, or with the dynamical content of the model, this discussion does not mess with the results obtained so far. However, we will not describe here all vacuum terms. We will present here only the most difficult vacuum terms which demand a careful analysis:
\begin{align}
\Sigma_V&=\int d^4x\left\{\alpha_1\bar{A}_\mu \bar{A}^\mu \bar{A}_\nu \bar{A}^\nu+\alpha_2m\bar{A}_\mu \bar{A}^\mu \bar{A}_\nu E^\nu +\alpha_3m^2\bar{A}_\mu \bar{A}^\mu E_\mu E^\mu +\alpha_4m^2\bar{A}_\mu \bar{A}_\nu E^\mu E^\nu+\right.\displaybreak[3]\nonumber\\
&+\left.\alpha_5m^3 \bar{A}_\mu E^\mu E_\nu E^\nu+\alpha_6m^4E_\mu E^\mu E_\nu E^\nu\right\}+s\int d^4x\left\{ \zeta\lambda_{\mu\nu\alpha}J^{\mu\beta\gamma}J^{\nu}_{\phantom{\nu} \beta\kappa}J_{\gamma}^{\phantom{\gamma}\kappa\alpha}+\right.\displaybreak[3]\nonumber\\
&+\left.\vartheta_1\bar{\kappa}_{\mu\nu\alpha\beta}\lambda^{\mu\rho\omega}J^{\nu}_{\phantom{\nu}\rho\sigma}J^{\alpha}_{\phantom{\alpha}\omega\delta}J^{\beta\sigma\delta}+\vartheta_2T_{\mu\nu\alpha\beta}^{\phantom{\mu\nu\alpha\beta}\theta\gamma}C_{\theta\gamma}\lambda^{\mu\rho\omega}J^{\nu}_{\phantom{\nu}\rho\sigma}J^{\alpha}_{\phantom{\alpha}\omega\delta}J^{\beta\sigma\delta}+\right.\displaybreak[3]\nonumber\\
&+\left.\vartheta_3T_{\mu\nu\alpha\beta}^{\phantom{\mu\nu\alpha\beta}\tau\zeta}\eta^{\gamma\xi}\bar{\kappa}_{\tau\gamma\zeta\xi}\lambda^{\mu\rho\omega}J^{\nu}_{\phantom{\nu}\rho\sigma}J^{\alpha}_{\phantom{\alpha}\omega\delta}J^{\beta\sigma\delta}\right\}\;,\displaybreak[3]\nonumber\\
&=\int d^4x\left\{\alpha_1\bar{A}_\mu \bar{A}^\mu \bar{A}_\nu \bar{A}^\nu+\alpha_2m\bar{A}_\mu \bar{A}^\mu \bar{A}_\nu E^\nu +\alpha_3m^2\bar{A}_\mu \bar{A}^\mu E_\mu E^\mu +\alpha_4m^2\bar{A}_\mu \bar{A}_\nu E^\mu E^\nu+\right.\displaybreak[3]\nonumber\\
&+\left.\alpha_5m^3 \bar{A}_\mu E^\mu E_\nu E^\nu+\alpha_6m^4E_\mu E^\mu E_\nu E^\nu\right\}+\int d^4x\left\{ \zeta J_{\mu\nu\alpha}J^{\mu\beta\gamma}J^{\nu}_{\phantom{\nu} \beta\kappa}J_{\gamma}^{\phantom{\gamma}\kappa\alpha}+\right.\displaybreak[3]\nonumber\\
&+\left.\vartheta_1\bar{\kappa}_{\mu\nu\alpha\beta}J^{\mu\rho\omega}J^{\nu}_{\phantom{\nu}\rho\sigma}J^{\alpha}_{\phantom{\alpha}\omega\delta}J^{\beta\sigma\delta}+\vartheta_2T_{\mu\nu\alpha\beta}^{\phantom{\mu\nu\alpha\beta}\theta\gamma}C_{\theta\gamma}J^{\mu\rho\omega}J^{\nu}_{\phantom{\nu}\rho\sigma}J^{\alpha}_{\phantom{\alpha}\omega\delta}J^{\beta\sigma\delta}+\right.\displaybreak[3]\nonumber\\
&+\left.\vartheta_3T_{\mu\nu\alpha\beta}^{\phantom{\mu\nu\alpha\beta}\tau\zeta}\eta^{\gamma\xi}\bar{\kappa}_{\tau\gamma\zeta\xi}J^{\mu\rho\omega}J^{\nu}_{\phantom{\nu}\rho\sigma}J^{\alpha}_{\phantom{\alpha}\omega\delta}J^{\beta\sigma\delta}\right\}\;.
\label{AV}
\end{align}
The terms that depend on the electron mass are introduced in order to guarantee the quantum stability of the vacuum action. This can be easily understood by fact that the sources $\bar{A}^{\mu}$ and $E^{\mu}$ suffer mix under quantum corrections. The same can be said about the sources $\bar{\kappa}_{\mu\nu\alpha\beta}$ and $C_{\nu\mu}$. 

At the physical limit of the sources \eqref{12}, the action \eqref{AV} reduces to 
\begin{eqnarray}
\Sigma_{Vphys}&=&\int d^4x\left\{\alpha_1a_\mu a^\mu a_\nu a^\nu+\alpha_2ma_\mu a^\mu a_\nu e^\nu +\alpha_3m^2a_\mu a^\mu e_\mu e^\mu +\alpha_4m^2a_\mu a_\nu e^\mu e^\nu+\right.\nonumber\\
&+&\left.\alpha_5m^3 a_\mu e^\mu e_\nu e^\nu+\alpha_6m^4e_\mu e^\mu e_\nu e^\nu+6\zeta v^4+(8\vartheta_3-2\vartheta_1)\kappa^{\alpha\mu\sigma}_{\phantom{\alpha\mu\sigma}\mu}v_\alpha v_\sigma v^2 +\right.\nonumber\\
&+&\left.8\vartheta_2c^{\alpha\sigma}v_\alpha v_\sigma v^2\right\}\;,
\label{AVP}
\end{eqnarray}
which shows the nontrivial vacuum of the model. Now we can proceed as in Sec.~\ref{MGC} and seek for the most general counterterm compatible with Ward identities shown at \eqref{22}:
\begin{eqnarray}
\Sigma_V^c&=&\int d^4x\left\{b_1\alpha_1\bar{A}_\mu \bar{A}^\mu \bar{A}_\nu \bar{A}^\nu+b_2\alpha_2m\bar{A}_\mu \bar{A}^\mu \bar{A}_\nu E^\nu +b_3\alpha_3m^2\bar{A}_\mu \bar{A}^\mu E_\mu E^\mu +\right.\nonumber\\
&+&\left.b_4\alpha_4m^2\bar{A}_\mu \bar{A}_\nu E^\mu E^\nu+b_5\alpha_5m^3 \bar{A}_\mu E^\mu E_\nu E^\nu+b_6\alpha_6m^4E_\mu E^\mu E_\nu E^\nu\right\}+\mathcal{S}_{\Sigma}\tilde{\Delta}^{(-1)}\;.
\end{eqnarray}
where
\begin{eqnarray}
\tilde{\Delta}^{(-1)}&=&\int d^4x\left\{b_7\zeta\lambda_{\mu\nu\alpha}J^{\mu\beta\gamma}J^{\nu}_{\phantom{\nu} \beta\kappa}J_{\gamma}^{\phantom{\gamma}\kappa\alpha}+b_8\vartheta_1\bar{\kappa}_{\mu\nu\alpha\beta}\lambda^{\mu\rho\omega}J^{\nu}_{\phantom{\nu}\rho\sigma}J^{\alpha}_{\phantom{\alpha}\omega\delta}J^{\beta\sigma\delta}+\right.\nonumber\\
&+&\left.b_9\vartheta_2T_{\mu\nu\alpha\beta}^{\phantom{\mu\nu\alpha\beta}\theta\gamma}C_{\theta\gamma}\lambda^{\mu\rho\omega}J^{\nu}_{\phantom{\nu}\rho\sigma}J^{\alpha}_{\phantom{\alpha}\omega\delta}J^{\beta\sigma\delta}+b_{10}\vartheta_3T_{\mu\nu\alpha\beta}^{\phantom{\mu\nu\alpha\beta}\tau\zeta}\eta^{\gamma\xi}\bar{\kappa}_{\tau\gamma\zeta\xi}\lambda^{\mu\rho\omega}J^{\nu}_{\phantom{\nu}\rho\sigma}J^{\alpha}_{\phantom{\alpha}\omega\delta}J^{\beta\sigma\delta}\right\}\;.\nonumber\\
\end{eqnarray}
It is then straightforward to show that the most general counterterm is 
\begin{eqnarray}
\Sigma_V^c&=&\int d^4x\left\{b_1\alpha_1\bar{A}_\mu \bar{A}^\mu \bar{A}_\nu \bar{A}^\nu+b_2\alpha_2m\bar{A}_\mu \bar{A}^\mu \bar{A}_\nu E^\nu +b_3\alpha_3m^2\bar{A}_\mu \bar{A}^\mu E_\mu E^\mu +\right.\nonumber\\
&+&\left.b_4\alpha_4m^2\bar{A}_\mu \bar{A}_\nu E^\mu E^\nu+b_5\alpha_5m^3 \bar{A}_\mu E^\mu E_\nu E^\nu+b_6\alpha_6m^4E_\mu E^\mu E_\nu E^\nu+\right.\nonumber\\
&+&b_7\zeta J_{\mu\nu\alpha}J^{\mu\beta\gamma}J^{\nu}_{\phantom{\nu}\beta\kappa}J_{\gamma}^{\phantom{\gamma}\kappa\alpha}+b_8\left.\vartheta_1\bar{\kappa}_{\mu\nu\alpha\beta}J^{\mu\rho\omega}J^{\nu}_{\phantom{\nu}\rho\sigma}J^{\alpha}_{\phantom{\alpha}\omega\delta}J^{\beta\sigma\delta}+\right.\nonumber\\
&+&\left.b_9\vartheta_2T_{\mu\nu\alpha\beta}^{\phantom{\mu\nu\alpha\beta}\theta\gamma}C_{\theta\gamma}J^{\mu\rho\omega}J^{\nu}_{\phantom{\nu}\rho\sigma}J^{\alpha}_{\phantom{\alpha}\omega\delta}J^{\beta\sigma\delta}+b_{10}\vartheta_3T_{\mu\nu\alpha\beta}^{\phantom{\mu\nu\alpha\beta}\tau\zeta}\eta^{\gamma\xi}\bar{\kappa}_{\tau\gamma\zeta\xi}J^{\mu\rho\omega}J^{\nu}_{\phantom{\nu}\rho\sigma}J^{\alpha}_{\phantom{\alpha}\omega\delta}J^{\beta\sigma\delta}\right\}\;.\nonumber\\
\end{eqnarray}
To finish the renormalizability of the vacuum term it is needed to check the stability of the vacuum action. Thus, we have to show that
\begin{eqnarray}
\Sigma_V(J, \xi)+\varepsilon\Sigma^c_V(J,\xi)&=&\Sigma_V^0(J_0, \xi_0)+\mathcal{O}(\varepsilon^2)\;,
\label{J}
\end{eqnarray}
where the bare parameters are defined as
\begin{eqnarray}
\xi_0&=&Z_{\xi}\xi\;,\;\;\;\;\;\;\;\;\xi \in \left\{\zeta,\alpha_i,\vartheta_j\right\}\;.
\label{P}
\end{eqnarray}
It is not difficult to achieve the following consistent expressions:
\begin{eqnarray}
Z_{\alpha_1}&=&1+\varepsilon\left(b_1-4a_{15}+4a_3\right)\;,\nonumber\\
Z_{\alpha_2}&=&1+\varepsilon\left(b_2-a_4-3a_{15}-a_{10}+5a_3-4a_{16}\frac{\alpha_1}{\alpha_2}\right)\;,\nonumber\\
Z_{\alpha_3}&=&1+\varepsilon\left(b_3-2a_4-2a_{15}-2a_{10}+6a_3-a_{16}\frac{\alpha_2}{\alpha_3}\right)\;,\nonumber\\
Z_{\alpha_4}&=&1+\varepsilon\left(b_4-2a_4-2a_{15}-2a_{10}+6a_3-2a_{16}\frac{\alpha_2}{\alpha_4}\right)\;,\nonumber\\
Z_{\alpha_5}&=&1+\varepsilon\left(b_5-3a_4-a_{15}-3a_{10}+7a_3-2a_{16}\left(\frac{\alpha_3+\alpha_4}{\alpha_5}\right)\right)\;,\nonumber\\
Z_{\alpha_6}&=&1+\varepsilon\left(b_6-4a_4+4a_3-a_{16}\frac{\alpha_5}{\alpha_6}\right)\;,\nonumber\\
Z_{\zeta}&=&1+\varepsilon\left(b_7+4a_0\right)\;,\nonumber\\
Z_{\vartheta_1}&=&1+\varepsilon\left(b_8-a_1+5a_0\right)\;,\nonumber\\
Z_{\vartheta_2}&=&1+\varepsilon\left(b_9-a_5-a_6+a_3+4a_0-a_2\left(\frac{\vartheta_1-4\vartheta_3}{\vartheta_2}\right)\right)\;,\nonumber\\
Z_{\vartheta_3}&=&1+\varepsilon\left(b_{10}-a_1+5a_0-a_{10}\frac{\alpha_2}{\alpha_3}\right)\;.
\label{PR}
\end{eqnarray}

The proof that all other possible pure source term is also renormalizable follows the same prescription.

\section{Discrete mappings of the sources}\label{DM}
\begin{table}[h]
\centering
\begin{tabular}{|c|c|c|c|c|c|c|c|}
	\hline 
 sources & $C$ & $P$ & $T$ & $CP$ & $CT$ & $PT$ & $CPT$ \\
	\hline 
$C_{00}, \bar{\kappa}_{0i0j}, C_{ij}, \bar{\kappa}_{ijkl}$ & $+$ & $+$ & $+$ & $+$ & $+$ & $+$ & $+$  \\ 
$M_5,C_{0i}, C_{i0}, \bar{\kappa}_{0ijk}$& $+$ & $-$ & $-$ & $-$ & $-$ & $+$ & $+$  \\ 
$B_i, G_{i0j}, G_{ij0}, J_{0ij},\lambda_{0ij}$& $+$ & $+$ & $-$ & $+$ & $-$ & $-$ & $-$  \\ 
$B_0, G_{i00}, G_{ijk}, J_{ijk}, \lambda_{ijk}$ &  $+$ & $-$ & $+$ & $-$ & $+$  & $-$ & $-$ \\ 
$\bar{A}_0, E_0, F_i$ &  $-$ & $+$ & $+$ & $-$ & $-$  & $+$ & $-$ \\ 
$\bar{A}_i, E_i, F_0$ &  $-$ & $-$ & $-$ & $+$ & $+$  & $+$ & $-$ \\ 
$H_{ij}, D_{0i}, D_{i0}$ &  $-$ & $+$ & $-$ & $-$ & $+$  & $-$ & $+$ \\ 
$H_{0i}, D_{00}, D_{ij}$ &  $-$ & $-$ & $+$ & $+$ & $-$  & $-$ & $+$ \\ 
\hline 
\end{tabular}
\caption{Discrete mappings of the sources.}
\label{table4}
\end{table}
The coupling between the local sources and the Dirac bilinears depend on behavior of sources and Dirac bilinears under discrete mappings. Besides of the quantum numbers shown in the Table \ref{table3}, the discrete mappings displayed in Table \ref{table4} will also select the allowed couplings. 

\section{Feynman rules}\label{FR}

In this appendix, we provide the Feynman rules used in Sect.~\ref{ONELOOP}. Instead of dealing with the direct rules that could be extracted from the action \eqref{0}, we opt to treat the breaking terms as insertions. Thus, the set of propagators are the usual QED propagators: For the electron 
\begin{equation*}
\begin{tikzpicture}[line width=0.8 pt, scale=1.3]
	\draw[fermion] (3.0,5.0)--(5.0,5.0);
	\node at (6.0,5.0) {\Large$=\frac{i(\gamma^{\mu}p_{\mu}+m)}{p^2-m^2}$};
\end{tikzpicture}
\end{equation*}
For the photon:
\begin{equation*}
 \begin{tikzpicture}[line width=0.8 pt, scale=1.3]
	\draw[vector] (3,.5)--(5,.5);
	\node at (2.8,.5) {$\mu$};
	\node at (5.2,.5) {$\nu$};
	\node at (6.0,.5) {\Large$=-\frac{i\eta_{\mu\nu}}{p^2}$};
\end{tikzpicture}
\end{equation*}
where $p_{\mu}$ is the particle momentum. 

The fermion-photon vertex is given by
\begin{equation*}
\begin{tikzpicture}[line width=0.8 pt, scale=1.3]
	\draw[fermion] (-40:1)--(0,0);
	\draw[fermionbar] (40:1)--(0,0);
	\draw[vector] (180:1)--(0,0);
	\node at (0:2.0) {$=-ie\gamma^{\mu}$};
\end{tikzpicture}
\end{equation*}

The Feynman rules for the Lorentz-violating QED terms are obtained by insertions of Lorentz-violating coefficients in fermion and photon propagators, namely
\begin{equation*}
\begin{tikzpicture}[line width=0.8 pt, scale=1.3]
	\draw[fermion] (3.0,8.0)--(4.0,8.0);
	\draw[fermion] (4.0,8.0)--(5.0,8.0);
	\node at (4.0,8.0) {\Large$\times$};
	\node at (5.93,8.0) {$\;=\;-iM$};
	\draw[fermion] (3.0,7.3)--(4.0,7.3);
	\draw[fermion] (4.0,7.3)--(5.0,7.3);
	\draw[fill=black] (4.0,7.3) circle (.050cm);
	\node at (5.95,7.3) {$\;=\;i\Gamma^{\mu}p_{\mu}$};
	\draw[vector] (3.0,6.6)--(5.0,6.6);
	\node at (3.9,6.6) {\Large$\times$};
	\node at (2.8,6.6) {$\mu$};
	\node at (5.2,6.6) {$\nu$};
	\node at (6.5,6.6) {$\;=\;-2ip^{\alpha}p^{\beta}\kappa_{\alpha\mu\beta\nu}$};
	\draw[vector] (3.0,5.9)--(5.0,5.9);
	\draw[fill=black] (3.95,5.9) circle (.050cm);
	\node at (2.8,5.9) {$\mu$};
	\node at (5.2,5.9) {$\nu$};
	\node at (6.3,5.9) {$\;=\;2v^{\alpha}\epsilon_{\alpha\mu\beta\nu}p^{\beta}$};
\end{tikzpicture}
\end{equation*}
There is also an insertion at the fermion-photon vertex given by
\begin{equation*}
\begin{tikzpicture}[line width=0.8 pt, scale=1.3]
	\draw[fermion] (-40:1)--(0,0);
	\draw[fermionbar] (40:1)--(0,0);
	\draw[vector] (180:1)--(0,0);
	\node at (0:2) {$=-ie\Gamma^{\mu}$};
	\draw[fill=black] (0:0) circle (.050cm);\;.
\end{tikzpicture}
\end{equation*}

\section*{Acknowledgements}

The Conselho Nacional de Desenvolvimento Cient\'{i}fico e Tecnol\'{o}gico\footnote{RFS is a PQ-2 level researcher under the program \emph{Produtividade em Pesquisa}, 308845/2012-9.} (CNPq-Brazil), The Coordena\c c\~ao de Aperfei\c coamento de Pessoal de N\'ivel Superior (CAPES) and the Pr\'o-Reitoria de Pesquisa, P\'os-Gradua\c c\~ao e Inova\c c\~ao (PROPPI-UFF) are acknowledge for financial support.

\end{document}